\documentclass[
 onecolumn,
 11pt,
 amsmath,amssymb,
 aps,
 floatfix,
 superscriptaddress,
 nofootinbib
 ]{revtex4-2}
\usepackage{microtype}
\usepackage{bm}
\usepackage{physics}
\usepackage{mathtools}
\usepackage{nicefrac}
\usepackage[dvipsnames]{xcolor}
\usepackage{graphicx}
\usepackage[sort&compress]{natbib}
\usepackage[hidelinks]{hyperref}
\hypersetup{pdftitle = {Noise-induced manifolds: reduction of slow--fast SDEs to switching processes}, pdfauthor = {Jeremy R. Worsfold, Patrick Ryan, and Richard G. Morris}}
\usepackage{cleveref}
\usepackage[version=4]{mhchem}
\usepackage{enumitem}
\usepackage{booktabs}

\usepackage{pifont}

\begin{document}

\author{Jeremy R. Worsfold}
\email{j.worsfold@unsw.edu.au}
\affiliation{
    School of Physics, UNSW, Sydney, NSW 2052, Australia.
}
\author{Patrick Ryan}
\affiliation{
    School of Physics, UNSW, Sydney, NSW 2052, Australia.
}
\author{Richard G. Morris}
\affiliation{
    School of Physics, UNSW, Sydney, NSW 2052, Australia.
}
\affiliation{
	EMBL Australia Node in Single Molecule Science, School of Biomedical Sciences, UNSW, Sydney, NSW 2052, Australia.
}
\affiliation{
	ARC Centre of Excellence for the Mathematical Analysis of Cellular Systems, UNSW Node, Sydney, NSW 2052, Australia.
}
\date{November 16, 2026}
\title{Noise-induced manifolds: reduction of slow--fast SDEs to switching processes}

\begin{abstract}
	When constituents copy one another, as in reaction networks, collective behaviour and opinion dynamics, the noise of copying vanishes at consensus and can make a system bistable.
	More generally this noise vanishes on whole manifolds, along which slower variables keep moving and set the switching rates, yet no general method derives these switching dynamics.
	Here we reduce slow--fast stochastic differential equations whose fast noise acts in a single direction to a switching diffusion or piecewise-deterministic Markov process (PDMP), using the absorption probabilities of the fast noise in place of an invariant measure, and derive a first-order correction for the finite duration of switches.
	For the two-species Togashi--Kaneko model the reduction recovers our earlier PDMP, and the correction captures how the stationary composition depends on the timescale separation.
	In a model of foraging ants with a depleting food source, the hazard of leaving a source rises through every stay, which makes the colony unlikely to leave a source soon after arriving there.
\end{abstract}

\maketitle

\section{Introduction}

Biological, ecological and social systems composed of many interacting constituents often admit descriptions in terms of a few collective variables.
When their dynamics separate into fast and slow processes, eliminating the fast degrees of freedom yields effective dynamics for the slow variables~\citep{pavliotis_multiscale_2008,bo_multiplescale_2017,parsons_dimension_2017}.

\begin{figure}[t]
	\centering
	\includegraphics[width=\linewidth]{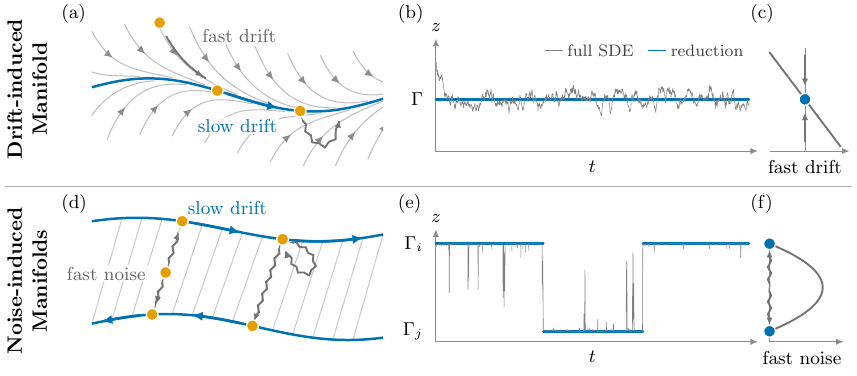}
	\caption{\textbf{Fast drift confines a system to one slow manifold, and fast noise makes it switch between several.}
	Fast moves are grey, manifolds and slow motion blue, and the state orange.
	(a) A strong fast drift carries every state onto one slow manifold $\Gamma$.
	(b) The fast coordinate $z$ of a sample path relaxes onto $\Gamma$ and fluctuates about it.
	(c) The fast drift vanishes only on $\Gamma$.
	(d) Strong noise acts along fibres and vanishes on two manifolds.
	The state is absorbed on either, moves along it, and later returns or switches to the other.
	(e) The sample path switches between $\Gamma_i$ and $\Gamma_j$.
	(f) The fast noise vanishes on both manifolds.}
	\label{fig:reduction-comparison}
\end{figure}

The canonical picture of these reductions is that of \cref{fig:reduction-comparison}a--c, in which a strong drift carries the fast variables quickly onto a slow manifold and the noise moves them only slightly off it.
In competing populations, for instance, density regulation returns the total size to its carrying capacity within a few generations, while weak selection and demographic noise change the composition over many, which makes the total a function of the composition~\citep{parsons_fixation_2007,lin_features_2012,rogers_demographic_2012,constable_stochastic_2013,chotibut_evolutionary_2015}.
Averaging over the fast equilibrium~\citep{kurtz_limit_1973,papanicolaou_probabilistic_1976,ball_asymptotic_2006} or projecting along the fast flow~\citep{katzenberger_solutions_1991,parsons_dimension_2017,ding_multiscale_2023} then gives the dynamics on this manifold.
In physics these constructions go by the name of adiabatic elimination, with the consistent threading being that the manifolds are a result of a fast deterministic drift.

In systems driven by copying, by contrast, the noise itself can create stable states, peaking the stationary distribution where it vanishes although the drift has a single fixed point~\citep{horsthemke_noiseinduced_1984,biancalani_noiseinduced_2012}.
When constituents copy one another, for instance when a molecule catalyses its own production, an animal follows the heading of a neighbour or an individual adopts another's opinion, copying that favours neither type cancels in the drift and \emph{adds up} in the noise.
In the system-size expansion this is the multiplicative noise of a stochastic differential equation (SDE), which vanishes at consensus, fixation or extinction and shrinks with the size of the system~\citep{kurtz_strong_1978,vankampen_stochastic_2007}.
Copying can nevertheless be frequent enough for its noise to dominate the slower processes, the regime usually described as a system below a critical size~\citep{togashi_transitions_2001,biancalani_noiseinduced_2014}.
Such noise-induced bistability has been found in autocatalytic reaction networks~\citep{togashi_transitions_2001,saito_motif_2016,jafarpour_noiseinduced_2015,sardanyes_noiseinduced_2018}, foraging ants and fish schools~\citep{biancalani_noiseinduced_2014,jhawar_noiseinduced_2020}, opinion dynamics~\citep{herrerias-azcue_consensus_2019,starnini_opinion_2026}, population genetics~\citep{bhat_fixation_2019,crosato_stochastic_2025} and gene regulation~\citep{dubnau_bistability_2006,raj_stochastic_2008,kohar_role_2018}.
Most analyses treat the simplest, and best understood, case, in which the states are two points, the peaks of a U-shaped distribution, and nothing else feeds back on the switching~\citep{horsthemke_noiseinduced_1984,houchmandzadeh_exact_2015,rosas_kramers_2016}.
More generally the noise vanishes on manifolds, along which slower variables keep moving and set the switching rates, as in the two-species Togashi--Kaneko (TK2) model, where exchange drives the composition to one of its pure states while immigration and removal keep changing the total~\citep{worsfold_stochastically_2025a}.
In earlier work we found that, when exchange is fast, the TK2 dynamics on these sets, which we call noise-induced manifolds (\cref{fig:reduction-comparison}d--f), is captured by a piecewise-deterministic Markov process (PDMP)~\citep{worsfold_stochastically_2025a}.
Yet noise-induced bistability is found well beyond TK2, and it has not been made clear when persistent flows along noise-induced manifolds should be expected in these other systems, nor how their switching dynamics can be derived in general.

PDMPs, and their counterparts with slow noise, switching diffusions, are effective models in their own right~\citep{faggionato_nonequilibrium_2009,zeiser_autocatalytic_2010,gander_stochastic_2007,visco_switching_2010,hufton_intrinsic_2016,hufton_model_2019}.
In most such models, however, the discrete states and the rates of switching between them are imposed, as an environment that changes at random~\citep{kepler_stochasticity_2001,bressloff_stochastic_2017}, or come from species that are discrete from the outset, such as a promoter that is on or off~\citep{crudu_convergence_2012,kang_separation_2013}.
On noise-induced manifolds the states and rates are set by the dynamics itself, and deriving them requires extending the multiscale machinery behind adiabatic elimination to fast dynamics with several absorbing states.
The quasi-steady-state approximation has been extended to fast variables that switch between metastable states~\citep{bruna_model_2014}, which the fast noise leaves again, whereas a noise-induced manifold, once reached, can be left only through the slow dynamics.

Here, by presuming the timescale separation that produces the bistability, we derive these switching processes by a formal asymptotic expansion for fast noise that acts in a single direction, along \emph{fibres} that leave the slow variables unchanged and each join two noise-induced manifolds.
We reduce slow--fast SDEs of this kind to a switching process in which the slow variables move along the occupied manifold, and every switching rate is the product of how fast the slow processes push the state off that manifold and how likely the fast noise is then to carry it to the other manifold.
A substitution rate in population genetics has the same form, the supply of mutants times their fixation probability~\citep{champagnat_microscopic_2006}.
Committor-weighted rates therefore have precedents, and what the reduction adds is the persistent feedback between the switching and the slow variables that keep moving along the manifolds, together with a first-order correction for the finite time that a switch takes, which changes the drift and rates and displaces the slow variables at the end of every switch.
Since the timescales of most systems are only moderately separated, this correction widens the range of systems to which the reduction applies.
Because the rates depend on the slow variables, the reduction predicts how long a system stays in a noise-induced state and how that time changes as the slow variables move, and its construction points towards systems with more than two such manifolds, such as many-opinion voter models, competing species and coupled chromatin domains.

\begin{figure}[t]
	\centering
	\includegraphics[width=\linewidth]{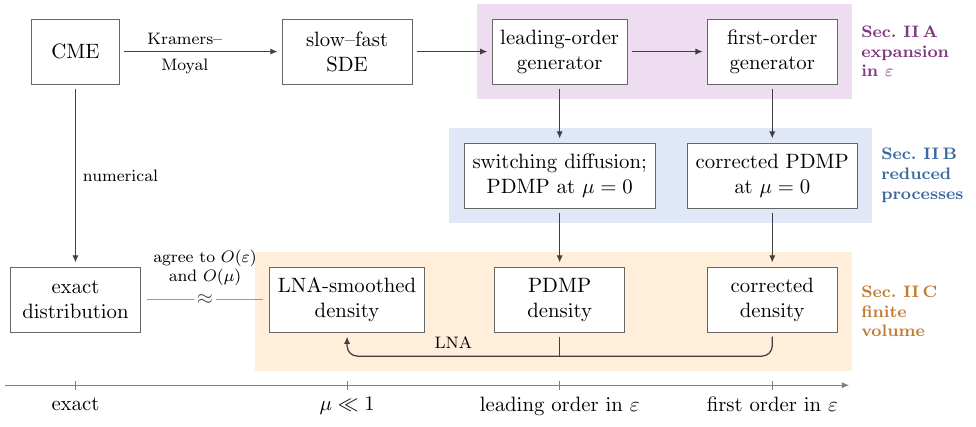}
	\caption{\textbf{The reduction takes a chemical master equation (CME) to a corrected switching process.}
	Each row is a step of \cref{sec:reduction}.
	The expansion in $\varepsilon$ gives general leading- and first-order generators (\cref{sec:singular-perturbation}), which \cref{sec:reset-pdmp-general} identifies as switching processes, and their stationary densities are smoothed to restore the finite-volume noise (\cref{sec:reaction-networks}).
	Columns are levels of approximation, where $\varepsilon$ is the timescale separation and $\mu = 1/V$ the inverse volume.}
	\label{fig:roadmap}
\end{figure}

\Cref{fig:roadmap} maps the steps of \cref{sec:reduction}, including the treatment of finite-volume noise in systems from chemical master equations (\cref{sec:reaction-networks}).
For TK2 (\cref{sec:tk2}) the reduction recovers our earlier PDMP~\citep{worsfold_stochastically_2025a}, and the correction captures how the stationary composition depends on the timescale separation.
In \cref{sec:ants} we apply the reduction to the recruitment model of foraging ants~\citep{kirman_ants_1993,biancalani_noiseinduced_2014}, in which the colony switches between two food sources, and let one of the sources deplete while it is exploited.
\Cref{sec:discussion} returns to systems with more manifolds and to the limits of the reduction.

\section{SDE reduction to noise-induced manifolds}
\label{sec:reduction}
In this section we seek a description that lives on the noise-induced manifolds alone, recording which manifold the system is on and where along it.
It should reproduce the statistics of the full dynamics as $\varepsilon \to 0$, with a systematic correction at small but finite $\varepsilon$.
We will find that the leading order needs only the probabilities of absorption on each manifold, and that the first-order correction needs in addition one ordinary differential equation along each fibre.
We consider an It\^o SDE on a domain $\Omega\subset\mathbb R^d$ whose noise has a fast part and may also have a slow component,
\begin{align}
	\dot x = a(x) + \varepsilon^{-\nicefrac{1}{2}}\sigma_f(x)\,\eta_f(t) + \sigma_s(x)\,\eta_s(t),
	\label{eq:setup-sde}
\end{align}
where the drift $a\colon\Omega\to\mathbb R^d$ and the noise amplitudes $\sigma_f\colon\Omega\to\mathbb R^{d\times m_f}$ and $\sigma_s\colon\Omega\to\mathbb R^{d\times m_s}$ are twice continuously differentiable in the interior of $\Omega$, $\eta_f$ and $\eta_s$ are independent Gaussian white noises in $\mathbb R^{m_f}$ and $\mathbb R^{m_s}$, and $0 < \varepsilon \ll 1$ is the ratio of the fast and slow timescales.
Its generator splits as
\begin{subequations}
\label{eq:setup-generator}
	\begin{align}
		\mathcal L &= \frac1\varepsilon\mathcal L_f + \mathcal L_s,
		\label{eq:setup-generator-L}\\
		\mathcal L_f &= B_f(x):\nabla^2,
		\label{eq:setup-generator-Lf}\\
		\mathcal L_s &= a(x)\cdot\nabla + B_s(x):\nabla^2,
		\label{eq:setup-generator-Ls}
	\end{align}
\end{subequations}
where $B_f=\tfrac12\sigma_f\sigma_f^{\mathsf T}$ and $B_s=\tfrac12\sigma_s\sigma_s^{\mathsf T}$, which we take to be twice continuously differentiable up to the manifolds, although the fast amplitude need not be, as for Wright--Fisher noise with amplitude proportional to $\sqrt{z(1 - z)}$.
The fast part has no drift, only noise, as for the neutral copying of the introduction (\cref{sec:reaction-networks}).
We work with this backward generator, which acts on observables.
In the Fokker--Planck picture, $\partial_t p = \mathcal{L}^* p$~\citep{gardiner_handbook_1985}, the fast relaxation produces Dirac measures on the manifolds whose derivatives are awkward to handle.
The forward equation \cref{eq:general-corrected-forward} follows once the reduced generator is known.

Throughout, we assume (A1) that the fast noise acts in a single direction, and thus $B_f$ has rank one wherever it does not vanish.
The reduced state space is the set of points where it vanishes.
Its connected components are the noise-induced manifolds $\Gamma_i$ of the introduction,
\begin{align}
	\textstyle\bigcup_i\Gamma_i = \{x : B_f(x) = 0\},
	\label{eq:fast-diffusion-manifolds}
\end{align}
which are absorbing for the fast dynamics alone.
Any quantity that the fast noise leaves unchanged is a slow variable.
We collect these in $y = \pi(x)$ and call the level sets of $\pi$, along which the fast noise acts, the \textit{fibres}.
Because the fast noise has rank one and no drift, it can conserve the $d - 1$ independent components of $\pi$ only along straight lines, and every fibre is therefore straight (\cref{app:index-derivation}).
We assume (A2) that the two ends of a fibre lie on two distinct manifolds, each of which it meets once and transversely.
From a point between the manifolds, the fast noise moves the state along its fibre, leaving the slow variables unchanged, until it reaches one of the two ends with probabilities given by the committors defined below (\cref{fig:reduction-comparison}d).
When the fast reactions exchange one of two species for the other, for example, the fast noise acts along $(1, -1)$ and conserves the total $\pi(x) = x_1 + x_2$, with the segments of constant total as fibres and the two axes as manifolds.

The reduction relies on two properties of the noise near the manifolds, stated in terms of the distance $\xi$ to $\Gamma_i$ along a fibre and a unit normal $n$ of $\Gamma_i$.
First (A3), the fast noise must reach the manifolds in finite time and be unable to leave them.
This holds when its variance normal to a manifold vanishes as
\begin{align}
	n\cdot B_f\,n \sim \xi^p, \qquad 1 \le p < 2.
	\label{eq:fast-exit-condition}
\end{align}
The Wright--Fisher, or voter-model, diffusion $z(1 - z)\partial_z^2$ of a fraction $z$ between two absorbing states has $p = 1$.
So do both of our examples, because the variance of a fast reaction vanishes in proportion to the amount of the species that runs out.
In Feller's classification~\citep{feller_diffusion_1954} the manifolds are then exit boundaries, and the fast process must end on one of them.
(Noise that vanishes faster, with $p \ge 2$, would never reach the manifolds.)
Second (A4), the slow noise must have no component normal to the manifolds,
\begin{align}
	\left.n\cdot B_s\,n\right|_{\Gamma_i} = 0 \quad \text{for every unit normal } n \text{ of } \Gamma_i,
	\label{eq:slow-tangential-condition}
\end{align}
so that on a manifold the slow noise moves the state only tangentially.
Because $B_s$ is positive semidefinite, this condition also removes its tangent--normal components.
Under these two conditions we need not supply a boundary condition at the manifolds, because the coefficients alone determine what happens there.
The slow diffusion itself may also be of order one.

\Cref{sec:singular-perturbation} expands the generator in $\varepsilon$, and \cref{sec:reset-pdmp-general} evaluates the expansion for the slow--fast SDE \cref{eq:setup-sde} and identifies it as a switching process (\cref{fig:roadmap}).
Reaction networks, whose finite-volume noise breaks (A4), are treated in \cref{sec:reaction-networks}.

\subsection{Singular perturbation of the reduced generator}
\label{sec:singular-perturbation}

We capture the effect of the fast dynamics on an observable through the committors $q_j(x)$, the probabilities that the fast noise started at $x$ is absorbed on $\Gamma_j$.
These are the fixation probabilities of population genetics, and they solve
\begin{align}
	\mathcal{L}_f q_j = 0
\end{align}
with $q_j=\delta_{ij}$ on $\Gamma_i$.
Because the fibres are straight, the fast generator along a fibre from $\Gamma_i$ to $\Gamma_j$ is $b\,\partial_\xi^2$ in the distance $\xi$ from $\Gamma_i$, where $B_f = b\,ee^{\mathsf T}$ for the unit vector $e$ along the fibre.
The committors are therefore affine, $q_j = \xi/\ell$ and $q_i = 1 - \xi/\ell$ for a fibre of length $\ell(y)$, whatever the form of $b$.
In the examples we use the committor $z = q_j$ itself as the coordinate along a fibre, as in the Wright--Fisher diffusion above.
A reduced observable \(\phi=(\phi_i)\), a function of the slow variables $y = \pi(x)$ on each manifold, is lifted to the full state space by
\begin{align}
	(H\phi)(x) = \sum_i q_i(x)\phi_i(\pi), \label{eq:Hphi-lift}
\end{align}
where $H$ is the harmonic extension of $\phi$ to the full state space~\citep{grosskinsky_dynamics_2013}.
Let \(R\) restrict a full observable to the manifolds, $(Rf)_i = \left.f\right|_{\Gamma_i}$, with $RH=I$, $\mathcal P\equiv HR$ and $\mathcal Q\equiv I - \mathcal P$.
In other words, $R$ reads off the value of an observable on each manifold, while $H$ extends values on the manifolds back to the full space by weighting them with the committors.
Hence $H\phi(x)$ is the expected value of $\phi$ once the fast noise has absorbed the state started at $x$ on a manifold.
The projection $\mathcal P$ replaces an observable by its value after instantaneous absorption, and $\mathcal Q$ keeps the part of the observable that changes during the fast excursion.
In general, calligraphic letters denote operators acting within the full or reduced space, and $H$ and $R$ map between them.

Our construction follows the standard multiscale expansion of the generator~\citep{pavliotis_multiscale_2008}, carried out to first order with a projection operator for a single slow manifold by Ding and Xing~\citep{ding_multiscale_2023} (\cref{fig:reduction-comparison}a).
Averaging when the fast process has several invariant measures has been studied in \citep{freidlin_diffusion_1993,freidlin_averaging_2021,goddard_study_2023}, and fast diffusions with absorbing faces arise in the inclusion process and multi-allele Moran models~\citep{grosskinsky_dynamics_2013}.
We extend the expansion to a fast process that is a degenerate diffusion absorbed on several manifolds, along which the slow variables keep moving.
We seek an approximately invariant lift and reduced generator satisfying
\begin{align}
	\mathcal L^\varepsilon H^\varepsilon = H^\varepsilon\overline{\mathcal L}^{\varepsilon} + O(\varepsilon^2),
\end{align}
with expansions
\begin{subequations}
	\begin{align}
		H^\varepsilon &= H + \varepsilon H_1 + \varepsilon^2 H_2 + O(\varepsilon^3),\\
		\overline{\mathcal L}^{\varepsilon} &= \overline{\mathcal L}_0 + \varepsilon\overline{\mathcal L}_1 + O(\varepsilon^2).
	\end{align}
\end{subequations}
That is, applying the full generator to a lifted observable must give, up to $O(\varepsilon^2)$, the lift of the reduced generator applied to the same observable.
The reduced dynamics starts once the state has been absorbed on a manifold, after a time of order $\varepsilon$.
For a state that starts at $x$ between the manifolds, the adjoint of the lift supplies the initial condition, which at leading order places the state on $\Gamma_i$ with probability $q_i(x)$ at the slow coordinate $\pi(x)$.

At $O(\varepsilon^{-1})$ the expansion requires $\mathcal{L}_fH=0$, which holds by the definition of $H$.
At $O(1)$ it gives
\begin{align}
	\mathcal{L}_f H_1 + \mathcal{L}_s H = H\overline{\mathcal{L}}_0.
\end{align}
Applying $R$, and using $RH=I$ and $R\mathcal{L}_f=0$, we obtain the leading-order reduced generator
\begin{align}
	\overline{\mathcal{L}}_0 = R\mathcal{L}_s H.
\end{align}

For the first-order correction we need the corrected lift $H_1$, which follows from substituting $\overline{\mathcal{L}}_0$ back into the $O(1)$ equation,
\begin{align}
	\mathcal{L}_f H_1 = HR\mathcal{L}_s H - \mathcal{L}_s H = -\mathcal Q\mathcal L_s H,
\end{align}
whose right-hand side is the part of the slow evolution that instantaneous absorption does not capture.
To invert $\mathcal L_f$ we define the fast potential operator $\mathcal G$ through
\begin{align}
	-\mathcal L_f\mathcal G = -\mathcal G\mathcal L_f = \mathcal Q
	\label{eq:fast-potential-definition}
\end{align}
together with $R\mathcal G=0$ and $\mathcal GH=0$, which make the correction vanish on the manifolds and fix it uniquely.
Probabilistically, $\mathcal G$ integrates an observable over the duration of a fast excursion, $\mathcal G\psi(x) = \mathbb E_x\int_0^\tau(\mathcal Q\psi)(X^f_t)\,\dd t$, where $X^f_t$ is the fast process started at $x$ and $\tau$ its absorption time~\citep{oksendal_stochastic_2003}.
In practice we find $u = \mathcal G \psi$ by solving the Poisson problem
\begin{align}
	-\mathcal L_f u = \mathcal Q \psi = \psi - \sum_i q_i\left.\psi\right|_{\Gamma_i}, \qquad \left.u\right|_{\Gamma_i} = 0,
	\label{eq:fibre-poisson}
\end{align}
on each fibre separately.
Here $\left.\psi\right|_{\Gamma_i}$ is evaluated where the fibre through $x$ meets $\Gamma_i$, and $y$ enters only as a parameter.
On each fibre this is a two-point boundary-value problem, and its solution has a finite derivative at the manifolds under the exit condition \cref{eq:fast-exit-condition}.
The corrected lift is therefore
\begin{align}
	H_1\phi = \mathcal G\mathcal Q\mathcal L_s H\phi = \mathcal G\mathcal L_s H\phi.
\end{align}

At $O(\varepsilon)$ the expansion gives
\begin{align}
	\mathcal{L}_f H_2 + \mathcal{L}_s H_1 = H\overline{\mathcal{L}}_1 + H_1\overline{\mathcal{L}}_0.
\end{align}
Applying $R$ again and using $RH_1=0$, which makes the solution unique, we have $R\mathcal{L}_s H_1 = \overline{\mathcal{L}}_1$.
With the expression for $H_1$, this becomes the first-order generator
\begin{align}
	\overline{\mathcal{L}}_1 = R\mathcal{L}_s\mathcal{G}\mathcal{L}_s H.
\end{align}

Read directly, the leading-order result $\overline{\mathcal L}_0 = R\mathcal L_sH$ applies the slow dynamics to an observable that has already been averaged over where the fast noise will send the state, and reads off the result on the manifolds.
The first-order result $\overline{\mathcal L}_1 = R\mathcal L_s\mathcal G\mathcal L_sH$ applies the slow dynamics once more, to what $\mathcal G$ has accumulated over the fast excursion.
In the next subsection we evaluate both for the SDE \cref{eq:setup-sde}.

\subsection{The reduced switching process}
\label{sec:reset-pdmp-general}

At leading order, a state pushed off a manifold is absorbed again instantaneously, on the same manifold or another.
We evaluate this for the slow--fast generator \cref{eq:setup-generator} by first applying the slow generator to the lift \cref{eq:Hphi-lift}, with the details in \cref{app:index-derivation}.
Writing $J_\pi = \partial\pi/\partial x$ for the Jacobian of the variables $y = \pi(x)$ conserved by the fast dynamics, the chain rule gives $\nabla\left[\phi_j(\pi(x))\right] = J_\pi^{\mathsf T}\nabla\phi_j$.
Applying this alongside the It\^o product rule, we obtain
\begin{equation}
	\mathcal L_s H\phi = \sum_j\Big[\phi_j\,\mathcal L_s q_j + \left(q_j\,\mathcal L_s\pi + 2\,J_\pi B_s\nabla q_j\right)\cdot\nabla\phi_j + q_j\,J_\pi B_sJ_\pi^{\mathsf T}:\nabla^2\phi_j\Big].
	\label{eq:LsH-expansion}
\end{equation}
On restricting with $R$, the cross term in \cref{eq:LsH-expansion} vanishes, because $q_j = \delta_{ij}$ on $\Gamma_i$ makes its gradient $\nabla q_j$ normal to the manifold, whereas the slow noise is tangential to it by \cref{eq:slow-tangential-condition}.
Moreover, $\sum_j q_j = 1$ implies $\sum_j\mathcal L_s q_j = 0$, which lets us write the terms in $\phi_j$ as gain and loss terms between manifolds.
The result is the backward generator of a switching diffusion
\begin{equation}
	(\overline{\mathcal L}_0\phi)_i(y) = v_i(y)\cdot\nabla\phi_i(y) + D_i(y):\nabla^2\phi_i(y) + \sum_{j\neq i}\lambda_{ij}(y)\left[\phi_j(y) - \phi_i(y)\right],
	\label{eq:general-leading-generator}
\end{equation}
with coefficients
\begin{subequations}
\label{eq:general-leading-coefficients}
	\begin{align}
		v_i(y) &= \left.\mathcal L_s\pi\right|_{\Gamma_i},
		\label{eq:general-leading-drift}\\
		D_i(y) &= \left.J_\pi\,B_s\,J_\pi^{\mathsf T}\right|_{\Gamma_i},
		\label{eq:general-leading-diffusion}\\
		\lambda_{ij}(y) &= \left.\mathcal L_s q_j\right|_{\Gamma_i}.
		\label{eq:general-leading-rates}
	\end{align}
\end{subequations}
\Cref{eq:general-leading-generator,eq:general-leading-coefficients} are the central result of this paper, a switching process whose states, flows and rates are read off from the full SDE.
In both examples the switching and its feedback with the slow variables already appear at this order, and the first-order correction below refines them.
In the language of stochastic trajectories, the slow variables $Y_t$ obey an It\^o SDE whose coefficients are those of the occupied manifold $I_t$,
\begin{align}
	\dd Y_t = v_{I_t}(Y_t)\,\dd t + \sqrt{2D_{I_t}(Y_t)}\,\dd W_t,
	\label{eq:general-leading-sde}
\end{align}
while the label $I_t$ jumps from $i$ to $j$ at rate $\lambda_{ij}(Y_t)$, and $Y_t$ is continuous across the jump.
A switch therefore changes the drift, diffusion and switching rates while leaving the slow variables where they are.
If, in addition, the slow noise has no component along the manifolds, then $D_i \equiv 0$ and between switches the slow variables follow the deterministic flow $\dot y = v_i(y)$ of the occupied manifold.
In this case the reduced process is a piecewise-deterministic Markov process (PDMP).

The switching rate $\lambda_{ij}$ is the rate at which the slow dynamics pushes the state off $\Gamma_i$, weighted by the probability that the fast noise then carries it to $\Gamma_j$.
On a flat manifold only the component of the slow drift normal to $\Gamma_i$ acts on the committor, through its normal gradient.
We require the rates to be nonnegative (A5), which there means that this normal drift points into the fibre or vanishes and keeps the slow dynamics from pushing the state out of the domain.
With two manifolds we write a single subscript for the source, so that $\lambda_\pm$ is the rate of leaving $\Gamma_\pm$, and likewise for the first-order coefficients below.

At first order, the finite duration of a fast excursion changes (i) the motion along the occupied manifold, (ii) the value of the slow coordinate at which the destination manifold is reached, and (iii) the probability of fixing on each destination manifold, and when the slow noise moves $\pi$ it also corrects the diffusion along the manifold and spreads the point of arrival.
To find these corrections, we recall that the corrected lift is $H_1\phi = \mathcal G\mathcal L_sH\phi$, with $\mathcal L_sH\phi$ given by \cref{eq:LsH-expansion}.
Because $\pi$ is conserved by the fast dynamics, functions of $\pi$ pass through $\mathcal G$ (\cref{eq:idx-G-commute}), and the fast potential therefore acts only on the coefficients of $\phi_j$, $\nabla\phi_j$ and $\nabla^2\phi_j$ in \cref{eq:LsH-expansion}.
For each coefficient, applying $\mathcal G$ means solving the fibre problem \cref{eq:fibre-poisson} with that coefficient as the source $\psi$.
Applying $R\mathcal L_s$ and using $\sum_j q_j = 1$, we find
\begin{equation}
	(\overline{\mathcal L}_1\phi)_i = v^{(1)}_i\cdot\nabla\phi_i + D^{(1)}_i:\nabla^2\phi_i + \sum_{j\neq i}\lambda^{(1)}_{ij}\left[\phi_j - \phi_i\right] + \sum_{j\neq i}\lambda_{ij}\left[r_{ij}\cdot\nabla\phi_j + Q_{ij}:\nabla^2\phi_j\right],
	\label{eq:general-first-order-generator}
\end{equation}
with coefficients for $j \neq i$
\begin{subequations}
\label{eq:general-first-order-coefficients}
	\begin{align}
		v_i^{(1)}(y) &= \left.\mathcal L_s\mathcal G\left[q_i\,\mathcal L_s\pi + 2\,J_\pi B_s\nabla q_i\right]\right|_{\Gamma_i},
		\label{eq:general-first-order-drift}\\
		D_i^{(1)}(y) &= \left.\mathcal L_s\mathcal G\left[q_i\,J_\pi B_sJ_\pi^{\mathsf T}\right]\right|_{\Gamma_i},
		\label{eq:general-first-order-diffusion}\\
		\lambda_{ij}^{(1)}(y) &= \left.\mathcal L_s\mathcal G\mathcal L_s q_j\right|_{\Gamma_i},
		\label{eq:general-first-order-rates}\\
		\lambda_{ij}(y)\,r_{ij}(y) &= \left.\mathcal L_s\mathcal G\left[q_j\,\mathcal L_s\pi + 2\,J_\pi B_s\nabla q_j\right]\right|_{\Gamma_i},
		\label{eq:general-first-order-reset}\\
		\lambda_{ij}(y)\,Q_{ij}(y) &= \left.\mathcal L_s\mathcal G\left[q_j\,J_\pi B_sJ_\pi^{\mathsf T}\right]\right|_{\Gamma_i}.
		\label{eq:general-first-order-reset-variance}
	\end{align}
\end{subequations}
Each coefficient applies $\mathcal L_s$ to a fibre solution that vanishes on every manifold, and on $\Gamma_i$ only the normal drift acting on its normal derivative survives.
The three corrections above enter through $v_i^{(1)}$, $\varepsilon r_{ij}$ and $\lambda_{ij}^{(1)}$, respectively, and the slow noise along $\pi$ adds the diffusion correction $D_i^{(1)}$ and the arrival spread $Q_{ij}$.
When the slow diffusion along $\pi$ vanishes identically, $J_\pi B_s J_\pi^{\mathsf T} \equiv 0$, we have $D_i^{(1)} = 0$ and $Q_{ij} = 0$.
To $O(\varepsilon)$ the corrected operator is then that of a PDMP whose switches are accompanied by resets.

Adding the first-order generator to the leading order, we obtain the corrected backward generator, which to $O(\varepsilon^2)$ takes the local form
\begin{equation}
	(\overline{\mathcal L}^{\varepsilon}\phi)_i = v_i^\varepsilon\cdot\nabla\phi_i + D_i^\varepsilon:\nabla^2\phi_i + \sum_{j\neq i}\lambda_{ij}^\varepsilon\left[\phi_j - \phi_i\right] + \varepsilon\sum_{j\neq i}\lambda_{ij}\left[r_{ij}\cdot\nabla\phi_j + Q_{ij}:\nabla^2\phi_j\right]
	\label{eq:general-corrected-backward}
\end{equation}
where $v_i^\varepsilon = v_i + \varepsilon v_i^{(1)}$, $D_i^\varepsilon = D_i + \varepsilon D_i^{(1)}$ and $\lambda_{ij}^\varepsilon = \lambda_{ij} + \varepsilon\lambda_{ij}^{(1)}$.
The off-diagonal derivatives in \cref{eq:general-corrected-backward} displace the slow variables when a switch lands, randomly when the $Q_{ij}$ term is present and, when $Q_{ij} = 0$, by the deterministic reset map
\begin{align}
	y \mapsto \Phi^\varepsilon_{ij}(y) = y + \varepsilon r_{ij}(y).
	\label{eq:general-reset-map}
\end{align}
Keeping the reset unexpanded defines a PDMP with nonlocal resets that can be simulated directly, which we call the corrected PDMP.
For calculations we use its local operator \cref{eq:general-corrected-backward}, which agrees with its generator to $O(\varepsilon)$.
Where we solve the local operator exactly, as in \cref{sec:tk2,sec:ants}, the solution contains higher powers of $\varepsilon$, of which only the first is controlled.
For the densities $\rho_i(y, t)$ on the manifolds, integrating by parts in $y$, with boundary terms that we assume vanish (A6), gives the forward equation
\begin{equation}
	\begin{aligned}
		\partial_t\rho_i ={}& -\nabla\cdot\left(v_i^\varepsilon\rho_i\right) + \nabla^2:\left(D_i^\varepsilon\rho_i\right) + \sum_{j\neq i}\left[\lambda_{ji}^\varepsilon\rho_j - \lambda_{ij}^\varepsilon\rho_i\right]\\
		&- \varepsilon\sum_{j\neq i}\nabla\cdot\left(\lambda_{ji}r_{ji}\rho_j\right) + \varepsilon\sum_{j\neq i}\nabla^2:\left(\lambda_{ji}Q_{ji}\rho_j\right),
	\end{aligned}
	\label{eq:general-corrected-forward}
\end{equation}
which at $\varepsilon = 0$ is the forward equation of the leading-order switching diffusion \cref{eq:general-leading-generator}.
The gain from $\Gamma_j$ is transported by $\varepsilon r_{ji}$ and spread by $\varepsilon Q_{ji}$ as it enters $\Gamma_i$.
The corrected PDMP, with its nonlocal resets, is a Markov process whenever the corrected rates are nonnegative, as they are for small enough $\varepsilon$ wherever $\lambda_{ij} > 0$, the diffusions $D_i^\varepsilon$ are positive semidefinite and the arrival points are drawn from a probability law with mean shift $\varepsilon r_{ij}$ and covariance $2\varepsilon Q_{ij}$.
Its local operator \cref{eq:general-corrected-backward} is a formal approximation to $O(\varepsilon)$, and because its derivatives act on the destination branch it need not preserve positivity, even with $Q_{ij}$ positive semidefinite.

In summary, the committors of a given model fix its leading-order process \cref{eq:general-leading-generator}, and the first-order correction takes three further steps: take the coefficients of $\mathcal L_sH\phi$ in \cref{eq:LsH-expansion} as sources, solve the fibre problems \cref{eq:fibre-poisson}, and take the normal derivatives of the solutions on each manifold.

\subsection{Reaction networks and finite-volume noise}
\label{sec:reaction-networks}

Reaction networks reach the slow--fast SDE \cref{eq:setup-sde} through the system-size expansion of the introduction, with $x$ a vector of concentrations in a volume $V$ and $\mu = 1/V$.
Their fast neutral reactions contribute the fast noise along a single stoichiometric direction, which makes the fibres parallel straight lines, and this noise vanishes where a reactant is absent.
We assume that it vanishes slowly enough to satisfy the exit condition \cref{eq:fast-exit-condition}.
The slow reactions supply the drift and a slow diffusion $B_s = O(\mu)$.

However, at finite volume the slow noise gains a component normal to the manifolds, violating the tangential condition \cref{eq:slow-tangential-condition}.
A reaction that triggers a switch creates a species absent on the manifold, and it therefore stays active there.
In the master equation the trigger is a single molecule, which the fast reactions then fix or lose, but the Kramers--Moyal truncation spreads it into a diffusion normal to the manifold.
Whether the continuum process is then absorbed, reflected or held for a while, as at a sticky boundary~\citep{harrison_sticky_1981}, is no longer fixed by the coefficients.
We therefore assume throughout that $\mu$ is small enough to take $\mu \to 0$ before reducing in $\varepsilon$.
The slow dynamics is then a pure drift and the tangential condition holds, and we restore the finite-volume fluctuations afterwards, keeping the $O(\varepsilon)$ and $O(\mu)$ corrections separately and dropping terms of order $\varepsilon\mu$.

To restore the fluctuations we follow \citet{hufton_intrinsic_2016} and use a linear noise approximation (LNA) about each branch flow.
On branch $i$ the slow coordinate obeys $\dd y = v_i\,\dd t + \sqrt{2\mu d_i}\,\dd W_t$, where $D_i = \mu d_i$ is the branch diffusion at finite volume.
Writing $y = \varphi + \sqrt{\mu}\,\zeta$ about the PDMP flow $\dot\varphi = v_i(\varphi)$ and linearising, we find that the fluctuation obeys $\dd\zeta = v_i'(\varphi)\,\zeta\,\dd t + \sqrt{2d_i(\varphi)}\,\dd W_t$, and its variance $w_i = \langle\zeta^2\rangle$ therefore changes at the rate $\dot w_i = 2v_i'w_i + 2d_i$.
Between switches $\varphi$ moves monotonically along the branch and thus by the chain rule $\dot w_i = v_iw_i'$, and the variance satisfies
\begin{align}
	v_iw_i' = 2d_i + 2v_i'w_i.
	\label{eq:lna-branch-variance}
\end{align}
On a given branch this equation determines $w_i$ up to a term $Cv_i^2$, with the constant $C$ fixed by the value of $w_i$ at the point where the system switched onto the branch.
Writing a tilde for stationary densities, we then approximate the finite-volume stationary distribution by smoothing the PDMP density $\tilde\rho_i$ on branch $i$ with the branchwise linear-noise kernel,
\begin{align}
	\tilde\rho^{(\mu)}_i(y) = \int\tilde\rho_i(\varphi)\,\mathcal N\left(y; \varphi, \mu w_i(\varphi)\right)\dd\varphi.
	\label{eq:lna-closure}
\end{align}
We validate this closure against the exact chain for TK2 in \cref{sec:tk2-finite-volume}.

\section{The two-species Togashi--Kaneko model}
\label{sec:tk2}

We first apply the reduction to the two-species Togashi--Kaneko (TK2) model, a minimal autocatalytic network in which the noise of a neutral exchange reaction creates bistability~\citep{togashi_transitions_2001,biancalani_noiseinduced_2012}.
With unequal removal rates, the composition switches between its two pure states while the total concentration grows on one and decays on the other (\cref{fig:tk2-leading}a)~\citep{worsfold_stochastically_2025a}.
Every step of the reduction can be done in closed form for TK2, whose slow total sets the switching rates, and the exact chain then tests the first-order correction.
The model consists of species $X_\pm$ with molecule numbers $n_\pm$ in a volume $V$, subject to immigration, removal and neutral autocatalytic exchange,
\begin{align}
	\emptyset\xrightarrow{\theta Vc_\pm}X_\pm, \quad X_\pm\xrightarrow{\theta k_\pm}\emptyset, \quad X_+ + X_-\xrightarrow{1}2X_\pm.
	\label{eq:tk2-reactions}
\end{align}
In molecule numbers the propensities are $\theta Vc_\pm$ for immigration, $\theta k_\pm n_\pm$ for removal and $n_+n_-/V$ for each exchange channel, where $\theta \ll 1$ makes immigration and removal slow.
We set $c_+ + c_- = 1$ and $k_+ + k_- = 2$ and measure time in units of $1/\theta$, in which immigration and removal act at rates of order one.
Neutral exchange fixes the composition in a time of order $V$.
The timescale separation $\varepsilon = \theta V$ is the ratio of this time to $1/\theta$, held fixed as $V \to \infty$.
We write $c_\pm = (1 \pm \beta)/2$ and $k_\pm = 1 \pm \kappa$, where $\beta$ is the influx asymmetry and $\kappa$ the removal-rate contrast.
Let \(x=(x_+,x_-)=\mu(n_+,n_-)\) denote the concentrations, with $\mu=1/V$.

Truncating the Kramers--Moyal expansion of the corresponding master equation at second order, we obtain the It\^o SDE~\citep{togashi_transitions_2001,biancalani_noiseinduced_2012,worsfold_stochastically_2025a}
\begin{equation}
	\dot x_\pm = c_\pm - k_\pm x_\pm \pm \varepsilon^{-\nicefrac12}\sqrt{2x_+x_-}\,\eta_f(t) + \sqrt{\mu\left(c_\pm + k_\pm x_\pm\right)}\,\eta_\pm(t),
	\label{eq:tk2-sde}
\end{equation}
with $\eta_f,\eta_+,\eta_-$ independent white noises.
Immigration and removal each change one species alone, leaving their noises independent of each other and of the other species.
The two exchange channels both move along $(1, -1)$, cancelling in the drift and adding in the noise.
They share the single noise $\eta_f$, which couples the species (its factor of $\mu$ is absorbed into $\varepsilon$).
\Cref{eq:tk2-sde} is of the slow--fast form \eqref{eq:setup-sde}, with generator $\mathcal L^\varepsilon = (1/\varepsilon)\mathcal L_f+\mathcal L_s$, where
\begin{subequations}
	\begin{align}
		\mathcal L_f &= x_+x_-\left(\partial_{x_+} - \partial_{x_-}\right)^2, \label{eq:tk2-fast-generator}\\
		\mathcal L_s &= \sum_\pm\left[(c_\pm - k_\pm x_\pm)\partial_{x_\pm} + \frac\mu2\left(c_\pm + k_\pm x_\pm\right)\partial_{x_\pm}^2\right],
	\end{align}
\end{subequations}
are the generators of the fast neutral autocatalytic diffusion and of slow immigration and removal, respectively.
Since the fast noise acts along $(1, -1)$, it conserves the total concentration $s = \pi(x) = x_+ + x_-$.
We work on $s > 0$, because immigration carries the state off the origin, where the two axes meet.
Each fibre $\{x_+ + x_- = s\}$ then ends on the two absorbing axes $\Gamma_\pm = \{x_\mp = 0, s > 0\}$ (\cref{fig:tk2-leading}b).
The committors $q_\pm = x_\pm/s$ are linear along each fibre and solve $\mathcal L_f q_\pm = 0$ subject to $q_\pm|_{\Gamma_\pm}=1$ and $q_\pm|_{\Gamma_\mp}=0$.
The lift, the fast projection and its complement are
\begin{subequations}
	\begin{align}
		(H\phi)(x) &= q_+\phi_+(s) + q_-\phi_-(s),\\
		(\mathcal P F)(x) &= q_+F(s, 0) + q_-F(0, s), \\ (\mathcal Q F)(x) &= F(x) - q_+F(s, 0) - q_-F(0, s),
	\end{align}
\end{subequations}
so that $\mathcal Q$ removes the linear interpolation of $F$ between its values at the two ends of the fibre.
At finite volume, however, immigration of the absent species adds slow noise normal to the absorbing axes, which violates the tangential condition \cref{eq:slow-tangential-condition}.
As in \cref{fig:roadmap}, we therefore assume $\mu = 0$ in the next two subsections, where the slow dynamics is a pure drift and the condition holds trivially.
In this limit we obtain a PDMP and its first-order correction, and we restore the finite-volume fluctuations in \cref{sec:tk2-finite-volume}.

\subsection{Leading order: a PDMP}

\begin{figure}[t]
	\centering
	\includegraphics[width=\linewidth]{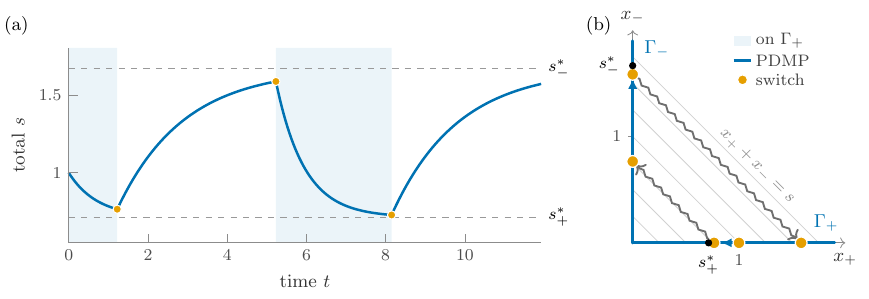}
	\caption{\textbf{At leading order the TK2 model reduces to a PDMP.}
	The total concentration flows towards a different fixed point on each manifold and sets the switching rate ($\kappa = 0.4$, $\beta = -0.2$).
	(a) The total concentration $s$ along a sample path of the leading-order PDMP (blue), shaded while the state is on $\Gamma_+ = \{x_- = 0\}$.
	On each manifold $s$ relaxes towards its fixed point $s^*_\pm$ (dashed), and a switch (orange) leaves it unchanged.
	(b) The same dynamics in the $(x_+, x_-)$ plane.
	The fast noise acts along the fibres $x_+ + x_- = s$ (grey) and vanishes on the axes, the manifolds $\Gamma_\pm$ (blue), along which $s$ flows towards the fixed points $s^*_\pm$ (black).
	The first two switches of the path in (a) run along fibres from one manifold to the other (grey wiggly arrows), from the state (orange) at $s = 1$.}
	\label{fig:tk2-leading}
\end{figure}

Substituting the pure drift of the slow generator and $J_\pi = (1, 1)$ into the leading-order coefficients \cref{eq:general-leading-coefficients}, we find the drifts and switching rates
\begin{align}
	v_\pm(s) &= \left.\mathcal L_s\pi\right|_{\Gamma_\pm} = 1 - k_\pm s, \\
	\lambda_\pm(s) &= \left.\mathcal L_s q_\mp\right|_{\Gamma_\pm} = \frac{c_\mp}{s}.
	\label{eq:tk2-leading-coefficients}
\end{align}
In units of $1/\theta$, the absent species immigrates at $Vc_\mp$ molecules per unit time, and one immigrant among $n = Vs$ fixes with probability $1/n$, giving $\lambda_\pm = c_\mp/s$.
Geometrically, $\lambda_\pm$ is the normal drift $c_\mp$ off $\Gamma_\pm$ times the normal gradient $\partial_{x_\mp}q_\mp = 1/s$.
We thereby recover the PDMP of \citet{worsfold_stochastically_2025a}, with reduced generator
\begin{align}
	(\overline{\mathcal L}_0\phi)_\pm = (1 - k_\pm s)\phi_\pm' + \frac{c_\mp}{s}\left[\phi_\mp - \phi_\pm\right]
	\label{eq:tk2-leading-generator}
\end{align}
and forward equation for the probability density $\rho_\pm=\rho_\pm(s,t)$ given by
\begin{align}
	\partial_t\rho_\pm = -\partial_s\left[v_\pm\rho_\pm\right] - \lambda_\pm\rho_\pm + \lambda_\mp\rho_\mp.
	\label{eq:tk2-pdmp-forward}
\end{align}
On each branch $s$ relaxes deterministically towards the fixed point $s^*_\pm = 1/k_\pm$ until a switch, at rate $c_\mp/s$, sends it towards the other, as shown in \cref{fig:tk2-leading}.
For $\kappa > 0$ the stationary density is supported between the two fixed points.
Setting the stationary currents to zero, we obtain the closed form
\begin{align}
	\tilde\rho(s) = \frac{\left(s - s^*_+\right)^{c_- - 1}\left(s^*_- - s\right)^{c_+ - 1}}{\Gamma(c_+)\,\Gamma(c_-)}
	\label{eq:tk2-leading-stationary}
\end{align}
for the total $\tilde\rho = \tilde\rho_+ + \tilde\rho_-$, where the normalisation is the Beta integral with $c_+ + c_- = 1$.
Given $s$, the probability of being on each manifold is the branch fraction
\begin{align}
	\frac{\tilde\rho_\pm(s)}{\tilde\rho(s)} = \frac{\pm(1 - k_\mp s)}{2\kappa s}.
	\label{eq:tk2-branch-fractions}
\end{align}

\subsection{First-order correction and reset representation}

With the leading order in hand, we now account for the time that a switch takes.
In the full model an excursion between the axes lasts a time of order $\varepsilon$, during which immigration and removal continue to act.
At $\mu = 0$ the slow noise is absent, and only two of the sources in the general first-order coefficients \cref{eq:general-first-order-coefficients} survive.
The rate $\mathcal L_s q_\pm$ at which the drift changes the committor sets the rate correction.
The committor-weighted rate of change of the total, $q_\pm\dot s$ with $\dot s = 1 - k_+x_+ - k_-x_-$, sets the drift correction and the reset.
Linear interpolation between the axes reproduces immigration and the common part of the removal exactly.
Hence only the removal-rate contrast enters the fibre problems, and all four reduce to a single Poisson problem along the fibre (\cref{app:tk2-first-order}).
Solving it, we find the coefficients
\begin{align}
	v^{(1)}_\pm = \pm\kappa c_\mp, \qquad \lambda^{(1)}_{\pm} = \pm\kappa\frac{c_\mp}{s}, \qquad r_{\pm} = \mp\kappa s,
	\label{eq:tk2-first-order-coefficients}
\end{align}
and $D^{(1)}_\pm = Q_\pm = 0$.
The drift correction, the rate correction and the reset flux $\lambda_\pm r_\pm$ all carry the factor $\kappa c_\mp$, because a minority immigration at rate $c_\mp$ initiates an excursion, and unequal removal acting on the mixed population during the excursion creates an additional imbalance proportional to $\kappa$.
The coefficients are, respectively, a correction to the drift on the source branch, a correction to the switching rate, and a displacement of $s$ on arrival at the destination branch.
The drift correction and displacement cancel in the mean, $v^{(1)}_\pm + \lambda_\pm r_\pm = 0$, as the first-order mean velocity \cref{eq:idx-mean-velocity} requires for $\dot s$ affine along the fibre.
In other words, the correction shifts part of the motion of $s$ from the source branch to the arrival after a switch.
Substituting \cref{eq:tk2-first-order-coefficients} into the corrected backward generator \cref{eq:general-corrected-backward}, we arrive at
\begin{equation}
	(\overline{\mathcal L}^{\varepsilon}\phi)_\pm = \left(1 - k_\pm s \pm \varepsilon\kappa c_\mp\right)\phi_\pm'(s) \mp \varepsilon\kappa c_\mp\phi_\mp'(s) + \frac{c_\mp}{s}(1 \pm \varepsilon\kappa)\left[\phi_\mp(s) - \phi_\pm(s)\right] + O(\varepsilon^2).
	\label{eq:corrected-local-generator}
\end{equation}
Since $Q_\pm = 0$ the displacement is deterministic, and the off-diagonal derivative is the first-order part of the reset map
\begin{align}
	\Phi_\pm^{\varepsilon}(s) = s + \varepsilon r_\pm(s) = (1 \mp \varepsilon\kappa)s,
\end{align}
which shrinks or stretches $s$ at each switch.
In the local form \cref{eq:general-corrected-backward} it appears as a transport term of the incoming density, and the forward equation \cref{eq:general-corrected-forward} becomes
\begin{equation}
	\partial_t\rho_\pm = -\partial_s\left[v_\pm^{\varepsilon}(s)\rho_\pm \pm \varepsilon\kappa c_\pm\rho_\mp\right] - \lambda_\pm^{\varepsilon}(s)\rho_\pm + \lambda_\mp^{\varepsilon}(s)\rho_\mp + O(\varepsilon^2),
	\label{eq:local-forward}
\end{equation}
with $v_\pm^{\varepsilon}(s) = 1 - k_\pm s \pm \varepsilon\kappa c_\mp$ and $\lambda_\pm^{\varepsilon}(s) = c_\mp(1 \pm \varepsilon\kappa)/s$.
Its stationary solution keeps the form of the leading-order density \cref{eq:tk2-leading-stationary},
\begin{align}
	\tilde\rho^\varepsilon(s) = \frac{\left(s - s^\varepsilon_+\right)^{a^\varepsilon_+ - 1}\left(s^\varepsilon_- - s\right)^{a^\varepsilon_- - 1}}{\Gamma(a^\varepsilon_+)\,\Gamma(a^\varepsilon_-)},
	\label{eq:tk2-corrected-stationary}
\end{align}
with endpoints and exponents
\begin{align}
	s^\varepsilon_\pm = \frac{2 - \varepsilon\kappa(\kappa + \beta) \mp \Delta}{2(1 - \kappa^2)}, \qquad a^\varepsilon_\pm = \frac12 \pm \frac{\varepsilon\kappa(\kappa + \beta) - 2\kappa\beta}{2\Delta},
	\label{eq:tk2-corrected-parameters}
\end{align}
where $\Delta = \sqrt{4\kappa^2 - 4\varepsilon\kappa^2(1 + \kappa\beta) + \varepsilon^2\kappa^2(\kappa + \beta)^2}$.
To first order in $\varepsilon$, $s^\varepsilon_\pm = (1 \pm \varepsilon\kappa c_\mp)/k_\pm$, the zeros of the corrected drifts $v^\varepsilon_\pm$, and $a^\varepsilon_\pm = c_\mp \pm \varepsilon\kappa(1 - \beta^2)/4$.
For $\kappa > 0$ the correction therefore moves weight towards $s^*_-$, the fixed point of the slower-removal branch, while the branch fractions \cref{eq:tk2-branch-fractions} are unchanged.

To assess the correction, we turn to the relative composition $m = 2z - 1 = (x_+ - x_-)/s$, which equals $\pm 1$ on $\Gamma_\pm$.
In the reduced process its stationary mean is $\langle m\rangle = P(+) - P(-)$, where $P(\pm) = \int\tilde\rho_\pm(s)\,\dd s$ is the stationary probability of being on $\Gamma_\pm$.
From the branch fractions \cref{eq:tk2-branch-fractions} we have $P(\pm) = \pm(\langle s^{-1}\rangle - k_\mp)/(2\kappa)$, and integrating $1/s$ against the corrected density \cref{eq:tk2-corrected-stationary} yields
\begin{align}
	\langle m\rangle = \frac{\langle s^{-1}\rangle - 1}{\kappa} = \frac{(s^\varepsilon_+)^{a^\varepsilon_+ - 1}(s^\varepsilon_-)^{-a^\varepsilon_+} - 1}{\kappa}.
	\label{eq:tk2-mean-composition}
\end{align}
At leading order this is $\langle m\rangle = [(1 + \kappa)^{c_+}(1 - \kappa)^{c_-} - 1]/\kappa$, independent of $\varepsilon$.
As $\kappa \to 0$ it tends to $\beta$, its value for the equal-removal model of \citet{bibbona_stationary_2020}.

\subsection{Finite volume: the PDMP smoothed by the LNA kernel}
\label{sec:tk2-finite-volume}

The PDMP densities above are supported between the fixed points $s^*_\pm$, whereas at finite volume the exact density of $s$ extends beyond them.
The cause is the randomness of individual immigration and removal events, which we removed by taking $\mu \to 0$ before reducing.
To restore it, we first identify where this finite-volume noise enters the reduced dynamics, and then smooth the PDMP density, the last step in \cref{fig:roadmap}.
The immigration and removal reactions contribute the slow diffusion $B_s = \tfrac{\mu}{2}\operatorname{diag}(c_+ + k_+x_+, c_- + k_-x_-)$.
Projected onto the occupied manifold, it yields the tangential diffusion $D_\pm = \left.J_\pi B_sJ_\pi^{\mathsf T}\right|_{\Gamma_\pm} = \tfrac{\mu}{2}(1 + k_\pm s)$.
Its normal component, $(B_s)_{\mp\mp} = \mu c_\mp/2$ on $\Gamma_\pm$, violates the tangential condition \cref{eq:slow-tangential-condition}, and we evaluate $\overline{\mathcal L}_0 = R\mathcal L_sH$ from \cref{eq:LsH-expansion} with it.
The normal component corrects the rate through the second derivative of the committor, and the cross term $2B_s\nabla q_j\cdot\nabla\pi$ survives on the manifold (\cref{app:tk2-finite-volume}).
At leading order in $\varepsilon$ and first order in $\mu$ the finite-volume reduced generator is
\begin{equation}
	(\overline{\mathcal L}_0^{(\mu)}\phi)_\pm = \frac{c_\mp}{s}\left(1 - \frac{\mu}{s}\right)\left[\phi_\mp - \phi_\pm\right] + \mu\frac{c_\mp}{s}\left[\phi_\mp' - \phi_\pm'\right] + (1 - k_\pm s)\phi_\pm' + \frac{\mu}{2}(1 + k_\pm s)\phi_\pm''.
	\label{eq:tk2-finite-volume-generator}
\end{equation}
The two switching terms combine into a switch evaluated one molecule up,
\begin{equation*}
	\frac{c_\mp}{s + \mu}\left[\phi_\mp(s + \mu) - \phi_\pm(s + \mu)\right].
\end{equation*}
The minority immigrant is counted in the total and fixes with probability $1/(n + 1) = \mu/(s + \mu)$.
With this probability, \cref{eq:tk2-finite-volume-generator} is the Kramers--Moyal expansion of the exact reduced chain, in which each immigration adds one molecule and a minority immigrant switches the branch.
The normal noise that violates \cref{eq:slow-tangential-condition} thus sets where the switch takes place.

For the LNA closure of \cref{sec:reaction-networks}, we note that the slow dynamics on either branch is an immigration--death process whose linear-noise variance equals its mean.
The LNA variance equation \cref{eq:lna-branch-variance} then has the solution $w_\pm(\varphi) = \varphi$ on both branches, the only choice for which a switch leaves the variance unchanged along with $s$, and the smoothed density \cref{eq:lna-closure} becomes
\begin{align}
	\tilde\rho^{(\mu)}_\pm(s) = \int\tilde\rho_\pm(\varphi)\,\mathcal N(s; \varphi, \mu\varphi)\,\dd\varphi.
	\label{eq:tk2-pdmp-lna}
\end{align}
In summary, our finite-volume approximation is the $\mu = 0$ PDMP density, with its closed form, smeared by the LNA kernel.
We use the branchwise LNA smoothing as the finite-volume closure in place of the corrections to the rates and resets in \cref{eq:tk2-finite-volume-generator}.
For the corrected curve we smooth $\tilde\rho^\varepsilon_\pm$ with the same variance $\mu\varphi$.
Because the kernel is identical on both branches, smoothing leaves the mean composition \cref{eq:tk2-mean-composition} unchanged.

\begin{figure}[t]
	\centering
	\includegraphics[width=\linewidth]{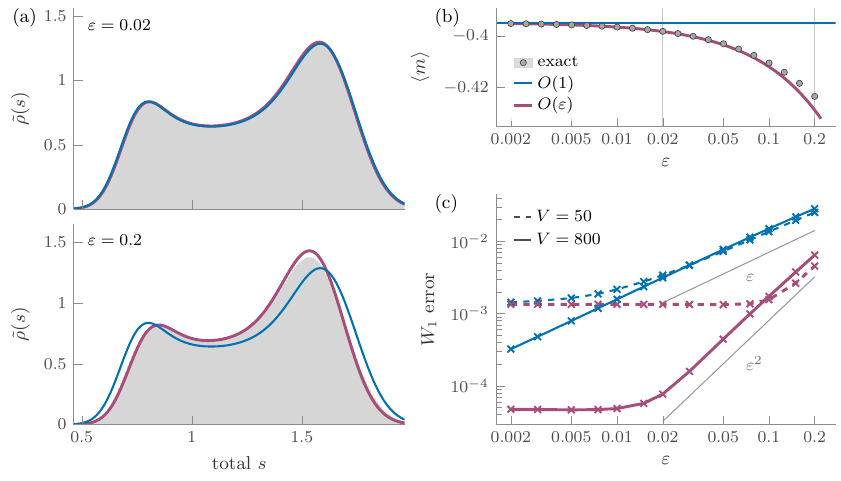}
	\caption{\textbf{The first-order correction extends the TK2 reduction to weaker timescale separation.}
	The stationary state changes with $\varepsilon$ and reduces the error from order $\varepsilon$ to order $\varepsilon^2$.
	The exact chain (grey) is compared with the leading-order (blue) and corrected (purple) PDMP ($\kappa = 0.4$, $\beta = -0.2$), the latter given by the stationary solution \cref{eq:tk2-corrected-stationary} of the local first-order operator.
	(a) Stationary density of $s$ at $V = 100$, with the PDMP densities smoothed by the LNA kernel \cref{eq:tk2-pdmp-lna}.
	(b) Mean relative composition $\langle m\rangle$ against $\varepsilon$, with grey lines at the values of $\varepsilon$ in (a).
	(c) Wasserstein distance $W_1$ between the exact and smoothed PDMP laws of $s$ against $\varepsilon$ at two volumes, with guides of slope 1 and 2.}
	\label{fig:tk2-correction}
\end{figure}

In \cref{fig:tk2-correction}(a,b) we compare the leading-order and corrected stationary solutions with the exact stationary distribution of the microscopic chain at $V = 100$.
We compute this distribution as the null vector of the generator of the chain, truncated by finite state projection at a cutoff that carries less than $10^{-8}$ of the stationary mass.
The corrected solution follows the exact mean composition $\langle m\rangle$ as it departs from the leading-order value, and tracks the exact density of $s$ more closely at larger $\varepsilon$.
The exact $\langle m\rangle$ does not depend on the volume to numerical precision, in agreement with the smoothed PDMP, whose mean composition the smoothing leaves unchanged.
We measure the error in the law of $s$ by the Wasserstein distance $W_1 = \int|F - F_{\mathrm{exact}}|\,\dd s$ between the cumulative distributions, which weights misplaced probability by how far it is moved.
This error combines the error of the reduction in $\varepsilon$ with the finite-volume error of the smoothing in $\mu$, and \cref{fig:tk2-correction}(c) shows it against $\varepsilon$ at two volumes.
At $V = 50$ the finite-volume error dominates and hides the convergence of the corrected PDMP in $\varepsilon$.
At $V = 800$ the error of the corrected PDMP falls like $\varepsilon^2$, as the expansion predicts, against $\varepsilon$ at leading order, until it meets the smaller finite-volume error.
For a given accuracy the correction therefore tolerates a much weaker separation of timescales, keeping the error at $V = 800$ below $10^{-3}$ up to $\varepsilon \approx 0.075$, where the leading order does so only up to $\varepsilon \approx 0.006$.
The remaining smoothing error falls with the volume, as shown in \cref{app:tk2-finite-volume}.

\section{Ant foraging with a depleting food source}
\label{sec:ants}

Ant colonies offered two identical food sources often exploit one of them far more heavily than the other~\citep{deneubourg_collective_1989,beckers_collective_1990}.
\citet{kirman_ants_1993} modelled this collective choice by recruitment, in which a forager converts a nestmate from the other source to its own, together with occasional changes of source by lone foragers.
Because recruitment converts one forager at a time and favours neither source, it is a form of neutral copying, whose noise vanishes once every forager is at one source.
When recruitment is much faster than the lone changes, most of the colony feeds at one source at a time and switches between the two at random, a noise-induced bistability whose mean switching time was obtained by \citet{biancalani_noiseinduced_2014}.

With sources that do not change, the reduced process leaves each source at a constant rate, and the stays are therefore exponential.
A food source is depleted while it is exploited and recovers when it is left.
For a depleting patch the marginal value theorem predicts that a forager should leave a patch once its rate of gain there falls to the average over the habitat~\citep{charnov_optimal_1976}.
We let one source deplete and regrow on a timescale comparable with a stay, keep the other steady, and assume that lone foragers leave a source faster as it becomes the worse of the two.
This assumption, motivated by patch depletion, lets the food left at the depleting source set the lone-switching rates.
Because it changes on the timescale of a stay, this food level is a slow variable of the reduction.
Its feedback on the switching is negative, since the hazard of leaving a source rises the longer the colony stays there.
In the reduced model this rising hazard delays the next switch after every switch (\cref{sec:ants-residence}), and at first order the switches that fail leave a trace of feeding at the abandoned source.

\begin{figure}[t]
	\centering
	\includegraphics[width=\linewidth]{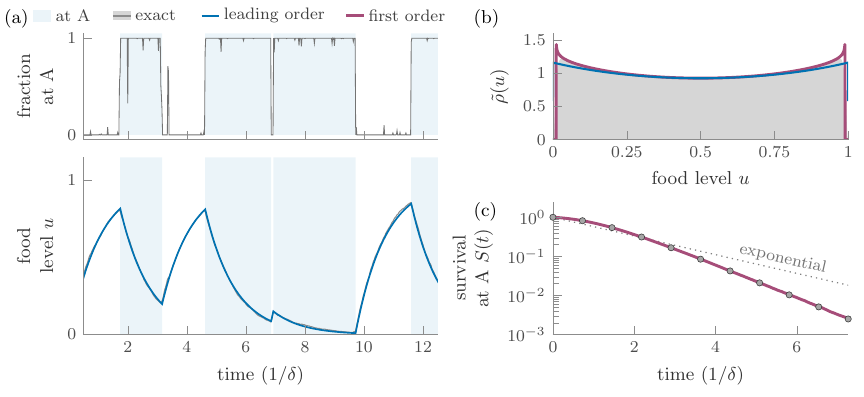}
	\caption{\textbf{Slow depletion of a food source makes the colony unlikely to switch again shortly after a switch in the foraging ant model.}
	The full system ($N = 200$, $\varepsilon = 0.1$, $\omega_0 = 0.02$, $\omega_1 = 0.2$, $\delta = 0.22$) is compared with the leading-order and corrected PDMPs.
	(a) A simulated path of the fraction of foragers at A, shaded during stays at A, and of the food level $u$ there, tracked by the leading-order flow driven by the same switches.
	(b) Stationary density of $u$, with the PDMP densities from the local operators.
	(c) Survival of stays at A in the exact chain and in simulated paths of the corrected PDMP with nonlocal resets, merged as in the text, against an exponential with the exact mean (dotted).
	The errors of both PDMPs are in \cref{fig:ants-errors}.}
	\label{fig:ants}
\end{figure}

\subsection{Model and reduction}

Each of $N$ foragers feeds at source A or source B, and we let $n$ count those at A.
One manifold has all $N$ at A, and the other has none.
Recruitment moves foragers between the sources, $n \to n \pm 1$ each at propensity $n(N - n)/\varepsilon$, as in the recruitment model~\citep{kirman_ants_1993,biancalani_noiseinduced_2014}.
A lone forager leaves B for A at rate $\omega_-$ and A for B at rate $\omega_+$.
The food level $u \in [0, 1]$ at source A falls at rate $\delta n/N$ as foragers feed there and rises at rate $\delta(1 - u)$ as A regrows.
Source B is a steady alternative although we could have also chosen this to be a depleting source.
Lone foragers leave A faster the less food is left there and leave B faster the more food A has,
\begin{align}
	\omega_+(u) = \omega_0 + \omega_1(1 - u)^2, \qquad \omega_-(u) = \omega_0 + \omega_1u^2,
	\label{eq:ants-switch-rates}
\end{align}
and with $\omega_1 = 0$ the sources are constant and the model is that of \citet{kirman_ants_1993}.

The fraction $z = n/N$ of foragers at A is fast.
Recruitment drives it with the fast noise $\mathcal L_f = z(1 - z)\partial_z^2$, of Wright--Fisher form, whose committor to the manifold at A is $q_+ = z$.
Since recruitment leaves the food level unchanged, $u$ is the slow coordinate $y$ of \cref{sec:reduction}, and the fibres are lines of constant food level.
The slow dynamics consists of the lone switches, which move $z$, and of feeding and regrowth, which move $u$,
\begin{equation}
	\dot z = (1 - z)\,\omega_-(u) - z\,\omega_+(u), \qquad \dot u = \delta(1 - u - z),
\end{equation}
where we assume $N$ large enough to drop the slow noise of lone switching, which is of order $1/N$.
The fast variance vanishes linearly at $z = 0$ and $z = 1$, which makes the manifolds at B and at A exit boundaries.
From the leading-order coefficients \cref{eq:general-leading-coefficients} we then find the reduced PDMP in the food level alone,
\begin{equation}
	\lambda_\pm(u) = \omega_\pm(u), \qquad v_+(u) = -\delta u, \qquad v_-(u) = \delta(1 - u),
	\label{eq:ants-leading}
\end{equation}
where $\lambda_+$ is the rate of leaving A and $\lambda_-$ the rate of leaving B.
Read as supply times fixation, the $N$ foragers at a source each leave it alone at rate $\omega_\pm$, and a lone defector converts the colony with probability $1/N$, cancelling the $N$ possible defectors.
Between switches the food level falls towards zero at A and recovers towards one at B (\cref{fig:ants}a).
At either source the hazard of leaving therefore rises through the stay towards $\lambda_+(0) = \lambda_-(1) = \omega_0 + \omega_1$.

At first order the lone-switching drift is affine in $z$ and drops out, as immigration does in TK2, leaving only the feeding term.
Solving the fibre problem for the feeding term (\cref{app:ants-coefficients}), we find from the general first-order coefficients \cref{eq:general-first-order-coefficients}
\begin{equation}
	v^{(1)}_\pm(u) = \pm\frac{\delta}{2}\,\lambda_\pm(u), \qquad \lambda^{(1)}_\pm = 0, \qquad r_\pm = \mp\frac{\delta}{2},
	\label{eq:ants-first-order}
\end{equation}
a deficit at A, a leak at B, and resets that cancel them in the mean, $v^{(1)}_\pm + \lambda_\pm r_\pm = 0$.
In a failed excursion from B, a lone forager reaches A and recruits a few others, who feed there briefly before the colony returns to B.
These excursions keep A slightly depleted while the colony feeds at B.
Conversely, failed excursions from A take foragers away from A for a while, and the resulting deficit slows the depletion of A while the colony is there.
The resets carry the feeding during the transit of a completed switch.
On a move to B they leave A more depleted by $\varepsilon\delta/2$ than an instantaneous switch would, because foragers not yet recruited away keep feeding at A.
On a move to A they leave it less depleted by $\varepsilon\delta/2$, because foragers not yet recruited to A do not feed there.

Again, the first-order terms cancel in the total current, which keeps the stationary branch densities at $\tilde\rho_+ = (1 - u)\tilde\rho$ and $\tilde\rho_- = u\tilde\rho$ at both orders.
The total density $\tilde\rho$ can then be calculated (\cref{app:ants-stationary}), and at leading order we find
\begin{align}
	\tilde\rho(u) \propto \left[u(1 - u)\right]^{a - 1}\exp\left[-\frac{\omega_1}{\delta}\,u(1 - u)\right], \qquad a = \frac{\lambda_-(1)}{\delta} = \frac{\lambda_+(0)}{\delta} = \frac{\omega_0 + \omega_1}{\delta},
	\label{eq:ants-stationary}
\end{align}
which with constant sources, $\omega_1 = 0$, is the Beta density with both exponents equal to $a$.
The exponent compares the rate of leaving a source once the food level has reached its fixed point with the regrowth rate.
For the parameters of \cref{fig:ants}, $a = 1$ and the density is a shallow U with its minimum at $u = 1/2$.
At first order the leak lowers the upper end of the support at $\varepsilon = 0.1$, to approximately $1 - \varepsilon\lambda_-(1)/2$, and the deficit raises the lower end symmetrically.

\subsection{Residence times at a source}
\label{sec:ants-residence}

We consider any two-manifold PDMP of \cref{sec:reduction}, with branches $\pm$, slow variable $y\in[y_{\min},y_{\max}]$, drifts $v_\pm(y)$, rates $\lambda_\pm(y)$ and stationary densities $\tilde\rho_\pm(y)$.
The survival function $S_\pm(t) = P(T_\pm > t)$ of the residence time $T_\pm$ on a branch averages the survival from each entry point $y$ over the stationary density $w_\pm(y)$ of entry points,
\begin{align}
	S_\pm(t) = \int_{y_{\min}}^{y_{\max}} w_\pm(y)G_\pm(t|y)\,\dd y.
	\label{eq:residencesurvival}
\end{align}
From an entry point the slow variable follows the flow $\dot\psi_\pm = v_\pm(\psi_\pm)$ with $\psi_\pm(0) = y$.
Along this flow switching is an inhomogeneous Poisson clock, with survival
\begin{align}
	G_\pm(t|y) = \exp\left(-\int_0^t \lambda_\pm\left(\psi_\pm(\tau)\right)\dd\tau\right).
	\label{eq:conditionalresidence}
\end{align}
A hazard that is constant along every branch trajectory, such as when the slow coordinate decouples from the switching, would make the residence times exponential.
With slow depletion, however, the rate of leaving a source changes during the stay, and its dependence on the food level at entry ties the length of a stay to the preceding stay at the other source.
As in the piecewise-deterministic treatment of gene activity intervals under feedback of \citet{pulkkinen_dynamics_2008}, the entry points are distributed as
\begin{align}
	w_\pm(y) = \frac{\lambda_\mp(y)\tilde\rho_\mp(y)}{J},
	\label{eq:entrypointdistr}
\end{align}
where $J=\int_{y_{\min}}^{y_{\max}}\lambda_\mp\tilde\rho_\mp\,\dd y$ is the total flux into branch $\pm$, equal at stationarity to the flux out of it.

For the ant model the flow at A from an entry food level $y$ is $\psi_+(\tau) = y\,e^{-\delta\tau}$, and from \cref{eq:conditionalresidence} we obtain
\begin{equation}
	G_+(t|y) = \exp\left[-(\omega_0 + \omega_1)t + \frac{2\omega_1y}{\delta}\left(1 - e^{-\delta t}\right) - \frac{\omega_1y^2}{2\delta}\left(1 - e^{-2\delta t}\right)\right],
	\label{eq:ants-conditional-survival}
\end{equation}
whose hazard $\lambda_+(\psi_+(t))$ rises from $\lambda_+(y)$ at entry towards $\omega_0 + \omega_1$.
Exchanging the sources yields $G_-(t|y) = G_+(t|1 - y)$.
Because the entry laws are also mirror images, $w_-(y) = w_+(1 - y)$, stays at B in the reduced process have the law of stays at A.
At first order the same formulas apply with the drifts and resets of \cref{eq:ants-first-order}, the flow carrying the leak or deficit and the entry point shifted by the reset.

In the reduced process a switch is a jump between manifolds, so that every stay has a sharp entry and exit.
In the exact chain a switch is a transit that takes a time of order $\varepsilon$, and as $\varepsilon$ grows the colony spends more of its time close to a manifold without being on it, in failed excursions and in the transits of switches.
We let a stay at A in the exact chain run from the first arrival at $n = N$ to the next arrival at $n = 0$, which assigns each failed excursion to the stay in which it starts.
The food steps make the exact chain finite, and we solve it directly with $M = 2000$, and $M = 8000$ for the errors against $\varepsilon$, adding to its state the label of the end visited last.
A reduced path can switch back before the first switch would have completed its transit, and the exact colony would then never have reached the other manifold, so that the pair of switches is a failed excursion.
To align the two descriptions, we simulate long paths of both PDMPs, the corrected one with its nonlocal resets \cref{eq:general-reset-map}, and merge each residence shorter than a transit with its neighbours into one stay.
Each transit time is drawn independently from the copying chain conditioned on fixation, with mean close to $\varepsilon$.
The merging lengthens the mean stay by an amount comparable with the first-order correction, and so it is needed for a comparison at first order.

The first-order correction reduces the error of the reduced process in both the density of the food level and the survival of a stay (\cref{fig:ants-errors}).
As for TK2, the Wasserstein error of the density falls from order $\varepsilon$ to order $\varepsilon^2$, down to the resolution set by the food steps of the exact chain (\cref{fig:ants-errors}a).
What error remains lies within about $0.03$ of either end, where the exact density rises smoothly from zero, whereas the corrected PDMP has a sharp edge, because the spread of the feeding during excursions enters only at second order.

The colony mostly arrives at a source when the rate of leaving it is low, and this rate rises through the stay.
A switch is therefore unlikely to be followed shortly by another, and a stay is unlikely to last much longer than the mean, which leaves fewer short and fewer long stays than an exponential with the same mean (\cref{fig:ants}c).
In this sense the stays are more regular than exponential, with a squared coefficient of variation of $0.47$ in the exact chain and in both merged PDMPs, against $1$ for an exponential.

\section{Discussion}
\label{sec:discussion}

Using the committors of the fast noise in place of an invariant measure, we reduced slow--fast systems with noise-induced manifolds to switching diffusions or PDMPs whose states and rates follow from the microscopic model.
At leading order a rate is the slow drift off a manifold times the committor gradient, which in both examples reads as the supply of immigrants or lone defectors times their fixation probability.
Read this way, a noise-induced switch is a fixation event, and the reduction carries the population-genetic picture of supply and fixation over to any system whose fast noise vanishes on manifolds.
In TK2 the first-order correction captures the shift in stationary composition as the timescale separation grows, and in the ant model the slow depletion suppresses both short and long stays at a source.
These effects come from the feedback between the switching and the slow variables, which a two-state chain with constant rates cannot represent.

Presuming the timescale separation lets the reduction drop the symmetry between species assumed in most analyses of noise-induced bistability, at the cost of the regime where noise and drift are comparable.
The first-order correction recovers part of this regime, which matters because the timescales of real systems are seldom cleanly separated, and in both examples it reaches a given accuracy at a timescale ratio about an order of magnitude larger than the leading order.
The two approaches are complementary, and the size of the first-order correction relative to the leading order indicates where the reduced description starts to fail.
For reaction networks, whose slow noise is not tangential to the manifolds (\cref{sec:reaction-networks}), we took $\mu \to 0$ first and restored the finite-volume broadening by LNA smoothing, which omits the mixed effects of the two limits.
Resolving the boundary from the master equation, where a switch starts from a single molecule, would remove the need for this ordering.
Both examples have two flat manifolds joined by fast noise acting in a single direction, and the first-order correction is a formal expansion tested against exact solutions without a proof of convergence.
On a curved manifold the slow noise along it adds a curvature term to the normal drift~\citep{hsu_stochastic_2002}.

The construction itself needs only the committors of the fast noise and the slow dynamics acting on them, which makes it a starting point for effective reductions of more complex systems.
Voter models with many opinions, networks of competing species and neighbouring chromatin domains have more than two noise-induced manifolds, and when the fast noise acts in several directions the committors solve a partial differential equation set by the geometry of the manifolds.
The reduced process is then a network of manifolds joined by switches, whose rates still take the form of a supply times a probability of fixation.
Biological systems in which fast copying creates states while slower processes feed back on them, such as kinase signalling~\citep{samoilov_stochastic_2005} and chromatin remodelling~\citep{bruno_epigenetic_2022,bruno_analysis_2024}, are natural candidates.
In these systems the reduction turns questions about how long a noise-induced state persists, and how that depends on the history of the slow variables, into questions about a low-dimensional switching process.
More broadly, wherever fast copying or alignment creates states while slower processes keep moving, the reduction offers a systematic route from the microscopic model to the switching process that describes it.

\section*{Data availability}
The code, data and figure sources used to produce the results in this paper are available at \href{https://github.com/jeremyworsfold/NI-manifolds-code}{github.com/jeremyworsfold/NI-manifolds-code}.
The version accompanying this preprint is tagged \href{https://github.com/jeremyworsfold/NI-manifolds-code/tree/preprint-v1}{\texttt{preprint-v1}}.

\appendix
\crefalias{section}{appendix}

\clearpage

\section{Details of the reduction}
\label{app:index-derivation}

Here we derive the steps of \cref{sec:reduction} that the main text states without proof.
Components of $x \in \Omega$ are indexed by $k, l, n$ and components of the slow coordinates $y = \pi(x)$ by $\alpha, \beta$, with repeated indices of either kind summed.
The branch indices $i, j$ are never summed implicitly.
We write $\partial_k = \partial/\partial x_k$ and $\partial_\alpha = \partial/\partial y_\alpha$, and $\psi(\pi)$ for a reduced function $\psi(y)$ evaluated at $y = \pi(x)$.

\subsection{Restriction to a manifold}

Both orders of the reduction require $\mathcal L_s$ applied to a product $c\,\psi(\pi)$ and restricted to $\Gamma_i$, where the coefficient $c(x)$ vanishes on $\Gamma_i$.
Using the product rule, we have
\begin{align}
	\mathcal L_s\left[c\,\psi(\pi)\right] = \psi\,\mathcal L_s c + c\,\mathcal L_s\left[\psi(\pi)\right] + 2\,(B_s)_{kl}\,\partial_k c\,\partial_l\pi_\alpha\,\partial_\alpha\psi.
	\label{eq:idx-product-reduced}
\end{align}
On $\Gamma_i$ the second part vanishes because $c = 0$ there.
To see that the third part also vanishes, we note that the gradient of $c$ is normal to $\Gamma_i$, because $c$ is constant there.
By \cref{eq:slow-tangential-condition} the slow noise has no normal component on $\Gamma_i$.
Since $B_s$ is positive semidefinite, zero variance $n_k(B_s)_{kl}n_l = 0$ along a normal $n$ implies $(B_s)_{kl}n_l = 0$, and hence
\begin{align}
	\left.(B_s)_{kl}\,\partial_l c\right|_{\Gamma_i} = 0.
	\label{eq:idx-cross-vanish}
\end{align}
Only the first part survives, and we find that the slow operator acts on the factor $c$ alone,
\begin{align}
	\left.c\right|_{\Gamma_i} = 0 \implies \left.\mathcal L_s\left[c\,\psi(\pi)\right]\right|_{\Gamma_i} = \psi\left.\mathcal L_s c\right|_{\Gamma_i}.
	\label{eq:idx-restriction}
\end{align}

At leading order we use this result on the lift.
Because the committors sum to one, we can write $q_i = 1 - \sum_{j\neq i}q_j$ and rearrange the lift as
\begin{align}
	H\phi = \phi_i(\pi) + \sum_{j\neq i}q_j\left[\phi_j - \phi_i\right](\pi),
	\label{eq:idx-lift-rearranged}
\end{align}
where every committor left in the sum vanishes on $\Gamma_i$.
Applying \cref{eq:idx-restriction} with $c = q_j$ to each term of the sum, we recover the switching terms of \cref{eq:general-leading-generator} with the rates \cref{eq:general-leading-rates}.

On a flat manifold the tangential second derivatives of $q_j$ vanish, and by \cref{eq:idx-cross-vanish} the slow diffusion does not act on the committor there.
The rate then reduces to $\lambda_{ij} = \left.a_n\,\partial_n q_j\right|_{\Gamma_i}$, the normal component of the slow drift times the normal gradient of the committor.
Since $q_j \geq 0$ takes its minimum on $\Gamma_i$, this gradient points into the fibre.
Hence the rate is nonnegative exactly when the slow drift does not point out of the domain, as assumption (A5) requires.
We also note that the drift $v_i = \left.\mathcal L_s\pi\right|_{\Gamma_i}$ contains the It\^o term $(B_s)_{kl}\,\partial_k\partial_l\pi_\alpha$ when $\pi$ is nonlinear in $x$ and the slow diffusion has a tangential component.
This term is the analogue of the noise-induced drift found by \citet{parsons_dimension_2017} for projection onto a single attracting manifold.

\subsection{Fibrewise action and straight fibres}

Here we show that the fast operators act on each fibre separately, with the slow coordinates as parameters, and that the fibres are straight.
The starting point is that the fast dynamics conserves the slow coordinates.
Applying It\^o's formula to $\pi_\alpha(X_t)$ under the fast process $dX_k = (\sigma_f)_{kr}\,dW_r$, we find that both its drift and its noise must vanish, which gives
\begin{align}
	\mathcal L_f\pi_\alpha = 0, \qquad (\sigma_f)_{kr}\,\partial_k\pi_\alpha = 0 \implies (B_f)_{kl}\,\partial_l\pi_\alpha = 0.
	\label{eq:idx-fast-conservation}
\end{align}
Next, we consider a product $c\,\chi(\pi)$ of a coefficient $c(x)$ and a reduced function $\chi(y)$.
In $\mathcal L_f$ applied to this product, every term except $\chi\,\mathcal L_f c$ contains one of the two quantities in \cref{eq:idx-fast-conservation}, and we are left with
\begin{align}
	\mathcal L_f\left[c\,\chi(\pi)\right] = \chi(\pi)\,\mathcal L_f c.
	\label{eq:idx-Lf-commute}
\end{align}
Multiplication by a function of $\pi$ thus commutes with $\mathcal L_f$.
It also commutes with $R$, $H$, $\mathcal P$ and $\mathcal Q$, since these act pointwise in $y$.
The fast potential inherits this property,
\begin{align}
	\mathcal G\left[c\,\chi(\pi)\right] = \chi(\pi)\,\mathcal G c.
	\label{eq:idx-G-commute}
\end{align}
To see this, we let $u = \chi(\pi)\,\mathcal G c$.
By \cref{eq:idx-Lf-commute} it satisfies $-\mathcal L_f u = \mathcal Q\left[c\,\chi(\pi)\right]$ and $Ru = 0$, the two conditions that define $\mathcal G\left[c\,\chi(\pi)\right]$ uniquely.

For the straightness of the fibres, we write $B_f = b\,ee^{\mathsf T}$, with $e$ a unit vector field and $b > 0$ in the interior of a fibre.
The conservation conditions \cref{eq:idx-fast-conservation} then read
\begin{align}
	e_k\,\partial_k\pi_\alpha = 0, \qquad e_ke_l\,\partial_k\partial_l\pi_\alpha = 0.
\end{align}
Differentiating the first condition along $e$ and subtracting the second, we find $(e\cdot\nabla e)\cdot\nabla\pi_\alpha = 0$.
The curvature $e\cdot\nabla e$ of the fibres is orthogonal to $e$, and by this identity it is also orthogonal to every $\nabla\pi_\alpha$.
Since the $d - 1$ gradients $\nabla\pi_\alpha$ span the directions orthogonal to $e$, the curvature vanishes.

Lastly, we check that the solution of the fibre problem \cref{eq:fibre-poisson} has a finite derivative at the manifolds.
In the distance $\xi$ along a fibre from one of its ends, the problem reads $-b\,u'' = \mathcal Q\psi$, where $\mathcal Q\psi$ vanishes linearly at the end and $b \sim \xi^p$.
Hence $u'' = O(\xi^{1 - p})$, and this is integrable under the exit condition $p < 2$ of \cref{eq:fast-exit-condition}.

\subsection{First-order generator and resets}

The corrected lift is $H_1\phi = \mathcal G\mathcal L_sH\phi$, since $\mathcal G\mathcal P = 0$.
Applying $\mathcal G$ to the expansion \cref{eq:LsH-expansion}, and using \cref{eq:idx-G-commute} to take $\phi_j$ and its derivatives outside, we obtain
\begin{align}
	H_1\phi = \sum_j\left[g_j\,\phi_j + h_{j\alpha}\,\partial_\alpha\phi_j + K_{j\alpha\beta}\,\partial_\alpha\partial_\beta\phi_j\right],
	\label{eq:idx-H1}
\end{align}
where we have introduced the correctors
\begin{subequations}
\label{eq:idx-correctors}
	\begin{align}
		g_j &= \mathcal G\mathcal L_s q_j,
		\label{eq:idx-corrector-g}\\
		h_{j\alpha} &= \mathcal G\left[q_j\,\mathcal L_s\pi_\alpha + 2\,(B_s)_{kl}\,\partial_k q_j\,\partial_l\pi_\alpha\right],
		\label{eq:idx-corrector-h}\\
		K_{j\alpha\beta} &= \mathcal G\left[q_j\,(B_s)_{kl}\,\partial_k\pi_\alpha\,\partial_l\pi_\beta\right].
		\label{eq:idx-corrector-K}
	\end{align}
\end{subequations}
Summing over branches and using $\sum_j q_j = 1$, we find the sum rules
\begin{align}
	\textstyle\sum_j g_j = \mathcal G\mathcal L_s 1 = 0, \qquad \sum_j h_{j\alpha} = \mathcal G\mathcal L_s\pi_\alpha, \qquad \sum_j K_{j\alpha\beta} = \mathcal G\left[(B_s)_{kl}\,\partial_k\pi_\alpha\,\partial_l\pi_\beta\right].
	\label{eq:idx-sum-rules}
\end{align}

Next, we apply $R\mathcal L_s$ to \cref{eq:idx-H1}.
Because $R\mathcal G = 0$, every corrector vanishes on every manifold.
The restriction \cref{eq:idx-restriction} then applies to each term, and we have
\begin{align}
	(\overline{\mathcal L}_1\phi)_i = (R\mathcal L_s H_1\phi)_i = \sum_j\left[\left.\mathcal L_s g_j\right|_{\Gamma_i}\phi_j + \left.\mathcal L_s h_{j\alpha}\right|_{\Gamma_i}\partial_\alpha\phi_j + \left.\mathcal L_s K_{j\alpha\beta}\right|_{\Gamma_i}\partial_\alpha\partial_\beta\phi_j\right].
	\label{eq:idx-first-order-sum}
\end{align}
The first sum rule in \cref{eq:idx-sum-rules} gives $g_i = -\sum_{j\neq i}g_j$.
Exactly as in \cref{eq:idx-lift-rearranged}, we can then write $\sum_j g_j\,\phi_j = \sum_{j\neq i}g_j\left[\phi_j - \phi_i\right]$, and the undifferentiated terms combine into switches at the rates $\lambda^{(1)}_{ij}$.
Separating $j = i$ from $j \neq i$ in the derivative terms, we arrive at \cref{eq:general-first-order-generator} with the coefficients \cref{eq:general-first-order-coefficients}.

The reset displacement $r_{ij}$ and the reset variance $Q_{ij}$ are defined by dividing by $\lambda_{ij}$, and they require $\lambda_{ij} > 0$ wherever the right-hand sides are nonzero.
Where this fails, the products $\lambda_{ij}r_{ij}$ and $\lambda_{ij}Q_{ij}$ remain well-defined coefficients of the local operator.

As a check, we apply \cref{eq:general-first-order-generator} to the observable $\phi_j(y) = y_\alpha$ on every branch.
For this observable $H\phi = \pi_\alpha$ and the switching terms vanish, and through the second sum rule we find
\begin{align}
	v^{(1)}_{i\alpha} + \sum_{j\neq i}\lambda_{ij}\,r_{ij\alpha} = \left.\mathcal L_s\mathcal G\mathcal L_s\pi_\alpha\right|_{\Gamma_i}.
	\label{eq:idx-mean-velocity}
\end{align}
The right-hand side is the first-order correction to the mean velocity of $y$ on branch $i$, which the reduced process splits between continuous drift along $\Gamma_i$ and jumps at switching events.

Finally, we show that the off-diagonal derivative terms of \cref{eq:general-corrected-backward} describe displacements at the switches.
Let a switch from branch $i$ to branch $j$ be accompanied by a random displacement $\Delta y$ of the slow coordinate with $\mathbb E[\Delta y_\alpha] = \varepsilon r_{ij\alpha} + O(\varepsilon^2)$, $\mathbb E[\Delta y_\alpha\Delta y_\beta] = 2\varepsilon Q_{ij\alpha\beta} + O(\varepsilon^2)$ and higher moments of order $o(\varepsilon)$.
Taylor expanding $\phi_j$ about $y$, we have
\begin{align}
	\mathbb E\left[\phi_j(y + \Delta y)\right] = \phi_j(y) + \varepsilon r_{ij\alpha}\,\partial_\alpha\phi_j(y) + \varepsilon Q_{ij\alpha\beta}\,\partial_\alpha\partial_\beta\phi_j(y) + o(\varepsilon),
\end{align}
and since $\lambda^\varepsilon_{ij} = \lambda_{ij} + O(\varepsilon)$, the jump term $\lambda^\varepsilon_{ij}\left(\mathbb E\left[\phi_j(y + \Delta y)\right] - \phi_i(y)\right)$ reproduces both the switching term and the off-diagonal derivative terms of \cref{eq:general-corrected-backward} to $O(\varepsilon)$.
This interpretation requires $Q_{ij}$ to be positive semidefinite.
When $Q_{ij} = 0$ the displacement is deterministic and equals the reset map \cref{eq:general-reset-map}.

\subsection{Fibres with Wright--Fisher noise}
\label{app:wf-fibres}

Both examples have two manifolds and the same fast noise.
Here we carry out the one fibre calculation behind their first-order coefficients.
We take the committor $z = q_+$ as the coordinate along a fibre, with $\Gamma_+$ at $z = 1$ and $\Gamma_-$ at $z = 0$, and a single slow coordinate $y$.
In these coordinates, and without slow noise, the two generators are
\begin{align}
	\mathcal L_f = z(1 - z)\,\partial_z^2, \qquad \mathcal L_s = \dot z\,\partial_z + \dot y\,\partial_y.
\end{align}
A function that vanishes on a manifold has no $y$ derivative there.
On such a function $\mathcal L_s$ therefore reduces to $\dot z\,\partial_z$.
From the leading-order rates \cref{eq:general-leading-rates} we have $\dot z = -\lambda_+$ at $z = 1$ and $\dot z = \lambda_-$ at $z = 0$.

Next, we note that the projection $\mathcal P$ interpolates a source linearly in $z$ between its values at the two ends.
Hence $\mathcal Q$ removes every part of the source that is affine in $z$.
In both examples what remains is a multiple $A(y)\,z(1 - z)$ of the fast variance.
For such a source the fibre problem \cref{eq:fibre-poisson} reads
\begin{align}
	-z(1 - z)\,\partial_z^2h = A\,z(1 - z), \qquad h(0) = h(1) = 0.
\end{align}
Dividing by $z(1 - z)$ and integrating twice, we find $h = \tfrac{A}{2}\,z(1 - z)$, with slopes $-A/2$ at $z = 1$ and $A/2$ at $z = 0$.
Combining these slopes with the values of $\dot z$ at the two ends, we arrive at
\begin{align}
	\mathcal G\left[A\,z(1 - z)\right] = \frac{A}{2}\,z(1 - z), \qquad \left.\mathcal L_s\mathcal G\left[A\,z(1 - z)\right]\right|_{\Gamma_\pm} = \frac{A}{2}\,\lambda_\pm.
	\label{eq:wf-corrector}
\end{align}
Every first-order coefficient of the examples is therefore the leading-order rate times half the prefactor of $z(1 - z)$ in its source.

\section{The TK2 model}
\label{app:tk2}

Here we evaluate the general coefficients of \cref{sec:reduction} for TK2, first at $\mu = 0$ for the first-order correction of \cref{sec:tk2} and then at finite volume for \cref{sec:tk2-finite-volume}.

\subsection{First-order coefficients}
\label{app:tk2-first-order}

At $\mu = 0$ every term of \cref{eq:general-first-order-coefficients} that contains $B_s$ vanishes, and the fast potential acts only on the sources $\mathcal L_s q_\pm$ and $q_\pm\mathcal L_s\pi = q_\pm\dot s$.
We evaluate them in the fibre coordinates $(s, z)$ of \cref{app:wf-fibres}, with $x_+ = sz$ and $x_- = s(1 - z)$.
Substituting these into immigration and removal, $\dot x_\pm = c_\pm - k_\pm x_\pm$, with $c_+ + c_- = 1$ and $k_+ - k_- = 2\kappa$, we find
\begin{align}
	\dot z = \frac{c_+ - z}{s} - 2\kappa\,z(1 - z), \qquad \dot s = 1 - k_-s - 2\kappa sz.
	\label{eq:tk2-fibre-dynamics}
\end{align}
At the two ends $\dot z$ returns the leading-order rates \cref{eq:tk2-leading-coefficients}, with $\dot z = -c_-/s = -\lambda_+$ at $z = 1$ and $\dot z = c_+/s = \lambda_-$ at $z = 0$.

Since $q_+ = z$ and $q_- = 1 - z$, the sources are $\mathcal L_s q_\pm = \pm\dot z$ and $q_\pm\dot s$.
Immigration and the common part of the removal are affine in $z$ and are removed by $\mathcal Q$.
Hence only the removal-rate contrast is left, and we have
\begin{align}
	\mathcal Q\mathcal L_s q_\pm = \mp2\kappa\,z(1 - z), \qquad \mathcal Q\left[q_\pm\dot s\right] = \pm2\kappa s\,z(1 - z).
	\label{eq:tk2-sources}
\end{align}
All four sources are multiples of $z(1 - z)$, and their fibre problems are therefore the single problem solved in \cref{eq:wf-corrector}.

Applying \cref{eq:wf-corrector} to each source with $\lambda_\pm = c_\mp/s$, we obtain the coefficients in \cref{tab:tk2-first-order}.
Dividing its last row by $\lambda_\pm$, we recover $r_\pm = \mp\kappa s$ and the coefficients \cref{eq:tk2-first-order-coefficients} of the main text.
Because $\dot s$ is affine in $z$, the sources $q_+\dot s$ and $q_-\dot s$ have opposite prefactors, which is the mean cancellation $v^{(1)}_\pm + \lambda_\pm r_\pm = 0$.
\begin{table}[h]
	\centering
	\begin{tabular}{@{}l@{\qquad}l@{\qquad}l@{\qquad}l@{}}
		\toprule
		Coefficient on $\Gamma_\pm$ & Source & $A/2$ & Value \\
		\midrule
		$\lambda^{(1)}_\pm$ & $\mathcal L_s q_\mp$ & $\pm\kappa$ & $\pm\kappa c_\mp/s$ \\
		$v^{(1)}_\pm$ & $q_\pm\dot s$ & $\pm\kappa s$ & $\pm\kappa c_\mp$ \\
		$\lambda_\pm r_\pm$ & $q_\mp\dot s$ & $\mp\kappa s$ & $\mp\kappa c_\mp$ \\
		\bottomrule
	\end{tabular}
	\caption{First-order coefficients of TK2 on $\Gamma_\pm$, each the rate $\lambda_\pm = c_\mp/s$ times half the prefactor $A$ of $z(1 - z)$ in its source, by \cref{eq:wf-corrector}.}
	\label{tab:tk2-first-order}
\end{table}

\subsection{Finite volume}
\label{app:tk2-finite-volume}

Here we derive the finite-volume terms of \cref{eq:tk2-finite-volume-generator}.
They come from the normal entry of $B_s$, which does not vanish on the manifolds because the absent species keeps immigrating.
We work on $\Gamma_+$, where $x_+ = s$ and $x_- = 0$, the normal direction is $x_-$ and $(B_s)_{--} = \mu c_-/2$.
The committor $q_- = x_-/(x_+ + x_-)$ vanishes all along $\Gamma_+$, and its $x_-$ derivatives there are
\begin{align}
	\partial_{x_-}q_- = \frac1s, \qquad \partial_{x_-}^2q_- = -\frac{2}{s^2}, \qquad \partial_{x_-}q_+ = -\frac1s,
\end{align}
where the last follows from $q_+ = 1 - q_-$.
These derivatives change two terms of \cref{eq:LsH-expansion}.
The first is the switching rate, which gains a contribution from the second derivative,
\begin{align}
	\lambda_{+-} = \left.\mathcal L_s q_-\right|_{\Gamma_+} = \frac{c_-}{s} + (B_s)_{--}\,\partial_{x_-}^2q_- = \frac{c_-}{s}\left(1 - \frac{\mu}{s}\right).
\end{align}
The second is the cross term, which survives through $(B_s)_{--}$ now that the tangential condition \cref{eq:slow-tangential-condition} fails,
\begin{align}
	\sum_j 2\,J_\pi B_s\nabla q_j\,\phi_j' = 2\,(B_s)_{--}\left(\partial_{x_-}q_-\,\phi_-' + \partial_{x_-}q_+\,\phi_+'\right) = \mu\frac{c_-}{s}\left(\phi_-' - \phi_+'\right).
\end{align}
Exchanging the labels, we obtain the terms on $\Gamma_-$, and together with the tangential diffusion $D_\pm = \tfrac{\mu}{2}(1 + k_\pm s)$ we arrive at \cref{eq:tk2-finite-volume-generator}.

\begin{figure}[t]
	\centering
	\includegraphics[width=0.6\linewidth]{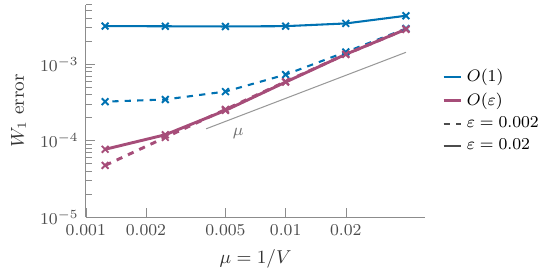}
	\caption{\textbf{The smoothing error of the corrected TK2 PDMP falls roughly in proportion to $\mu = 1/V$.}
	Wasserstein distance $W_1$ between the exact and smoothed PDMP laws of $s$ against $\mu$, at two values of $\varepsilon$ with the parameters of \cref{fig:tk2-correction}, with a guide of slope 1.}
	\label{fig:tk2-volume}
\end{figure}

In \cref{sec:tk2} we use the LNA smoothing \cref{eq:tk2-pdmp-lna} in place of this generator, and \cref{fig:tk2-volume} shows how its error depends on the volume.
At $\varepsilon = 0.02$ the error of the leading order is set by the reduction and barely changes with $\mu$, while the error of the corrected PDMP falls roughly like $\mu^{1.1}$.

\section{The ant foraging model}
\label{app:ants}

Here we derive the coefficients and the stationary density of the reduced ant model of \cref{sec:ants}.

\subsection{Reduced coefficients}
\label{app:ants-coefficients}

The manifold at A is $z = 1$ and the manifold at B is $z = 0$, with committors $q_+ = z$ and $q_- = 1 - z$.
With $\pi = u$ and no slow noise, the slow operator is
\begin{align}
	\mathcal L_s = \dot z\,\partial_z + \dot u\,\partial_u, \qquad \dot z = (1 - z)\,\omega_-(u) - z\,\omega_+(u), \qquad \dot u = \delta(1 - u - z).
\end{align}
A function that vanishes on a manifold has no $u$ derivative there, and on that manifold $\mathcal L_s$ reduces to $\dot z\,\partial_z$, with $\dot z = -\omega_+$ at $z = 1$ and $\dot z = \omega_-$ at $z = 0$.
At leading order the rates \cref{eq:general-leading-rates} are $\lambda_+ = \left.\dot z\,\partial_z q_-\right|_{z = 1} = \omega_+$ and $\lambda_- = \left.\dot z\,\partial_z q_+\right|_{z = 0} = \omega_-$.
The drifts are $v_+ = \dot u|_{z = 1}$ and $v_- = \dot u|_{z = 0}$, and together these give \cref{eq:ants-leading}.

At first order we use the fibre calculation of \cref{app:wf-fibres}, with the sources $\mathcal L_s q_\pm = \pm\dot z$ and $q_\pm\dot u$.
The lone-switching drift $\dot z$ is affine in $z$ and is removed by $\mathcal Q$, which gives $\lambda^{(1)}_\pm = 0$.
The feeding source $q_+\dot u = \delta z(1 - u - z)$ equals $-\delta u$ at $z = 1$ and vanishes at $z = 0$.
Subtracting its linear interpolation $-\delta uz$, we have
\begin{align}
	\mathcal Q\left[q_\pm\dot u\right] = \pm\delta\,z(1 - z),
	\label{eq:ants-sources}
\end{align}
where the lower sign follows from $q_-\dot u = \dot u - q_+\dot u$ and $\mathcal Q\dot u = 0$.

Applying \cref{eq:wf-corrector} to these sources with $\lambda_\pm = \omega_\pm$, we obtain the coefficients in \cref{tab:ants-first-order}.
Dividing its last row by $\lambda_\pm$, we recover $r_\pm = \mp\delta/2$ and the coefficients \cref{eq:ants-first-order} of the main text.
The opposite prefactors of $q_+\dot u$ and $q_-\dot u$ make the drift and reset corrections cancel in the mean, $v^{(1)}_\pm + \lambda_\pm r_\pm = 0$.
\begin{table}[h]
	\centering
	\begin{tabular}{@{}l@{\qquad}l@{\qquad}l@{\qquad}l@{}}
		\toprule
		Coefficient on $\Gamma_\pm$ & Source & $A/2$ & Value \\
		\midrule
		$\lambda^{(1)}_\pm$ & $\mathcal L_s q_\mp$ & $0$ & $0$ \\
		$v^{(1)}_\pm$ & $q_\pm\dot u$ & $\pm\delta/2$ & $\pm\tfrac{\delta}{2}\,\omega_\pm$ \\
		$\lambda_\pm r_\pm$ & $q_\mp\dot u$ & $\mp\delta/2$ & $\mp\tfrac{\delta}{2}\,\omega_\pm$ \\
		\bottomrule
	\end{tabular}
	\caption{First-order coefficients of the ant foraging model on $\Gamma_\pm$, each the rate $\lambda_\pm = \omega_\pm$ times half the prefactor $A$ of $z(1 - z)$ in its source, by \cref{eq:wf-corrector}.}
	\label{tab:ants-first-order}
\end{table}

\subsection{Stationary density}
\label{app:ants-stationary}

At leading order the branch densities obey
\begin{align}
	\partial_t\rho_\pm = -\partial_u\left(v_\pm\rho_\pm\right) - \lambda_\pm\rho_\pm + \lambda_\mp\rho_\mp.
\end{align}
Summing the two branches at stationarity, the total current $v_+\tilde\rho_+ + v_-\tilde\rho_-$ is constant in $u$, and it vanishes because no probability leaves $[0, 1]$.
With $v_+ = -\delta u$ and $v_- = \delta(1 - u)$, this fixes the branch densities as $\tilde\rho_+ = (1 - u)\tilde\rho$ and $\tilde\rho_- = u\tilde\rho$.
Substituting these into the stationary equation for the $+$ branch, we have
\begin{align}
	\delta\,\partial_u\left[u(1 - u)\tilde\rho\right] = \left[(1 - u)\,\omega_+(u) - u\,\omega_-(u)\right]\tilde\rho.
	\label{eq:ants-stationary-ode}
\end{align}
In terms of $F = u(1 - u)\tilde\rho$, this equation separates as
\begin{align}
	\frac{F'}{F} = \frac{1}{\delta}\left[\frac{\omega_+(u)}{u} - \frac{\omega_-(u)}{1 - u}\right] = a\left(\frac{1}{u} - \frac{1}{1 - u}\right) - \frac{\omega_1}{\delta}(1 - 2u),
\end{align}
where we have used $\omega_+(u) = \omega_0 + \omega_1 - \omega_1u(2 - u)$ and $\omega_-(u) = \omega_0 + \omega_1 - \omega_1(1 - u)(1 + u)$.
Integrating once and dividing by $u(1 - u)$, we recover \cref{eq:ants-stationary}.

At first order we use the local forward equation \cref{eq:general-corrected-forward} with $\lambda^{(1)}_\pm = 0$ and $Q_\pm = 0$.
Summed over the branches, the first-order terms add $\varepsilon\left(v^{(1)}_\pm + \lambda_\pm r_\pm\right)\tilde\rho_\pm$ to the total current, which vanishes by the mean cancellation.
The branch densities therefore keep the form $\tilde\rho_+ = (1 - u)\tilde\rho$ and $\tilde\rho_- = u\tilde\rho$.
The current on the $+$ branch now includes the deficit and the reset of arrivals from B, and \cref{eq:ants-stationary-ode} becomes
\begin{align}
	\delta\,\partial_u\left[\Theta_\varepsilon(u)\tilde\rho\right] = \left[(1 - u)\,\omega_+(u) - u\,\omega_-(u)\right]\tilde\rho,
\end{align}
with
\begin{align}
	\Theta_\varepsilon(u) = u(1 - u) - \frac{\varepsilon}{2}\left[(1 - u)\,\omega_+(u) + u\,\omega_-(u)\right].
\end{align}
Integrating this equation then gives
\begin{align}
	\tilde\rho^\varepsilon(u) \propto \frac{1}{\Theta_\varepsilon(u)}\exp\left[\frac{1}{\delta}\int^u\frac{(1 - \varphi)\,\omega_+(\varphi) - \varphi\,\omega_-(\varphi)}{\Theta_\varepsilon(\varphi)}\,\dd\varphi\right],
\end{align}
supported between the two zeros of $\Theta_\varepsilon$ in $(0, 1)$.
The upper zero lies close to $1 - \varepsilon\omega_-(1)/2$, where the leak balances the regrowth of A while the colony is at B, and the lower zero is its mirror image.

\subsection{Errors of the reduced process}
\label{app:ants-errors}

\begin{figure}[t]
	\centering
	\includegraphics[width=0.8\linewidth]{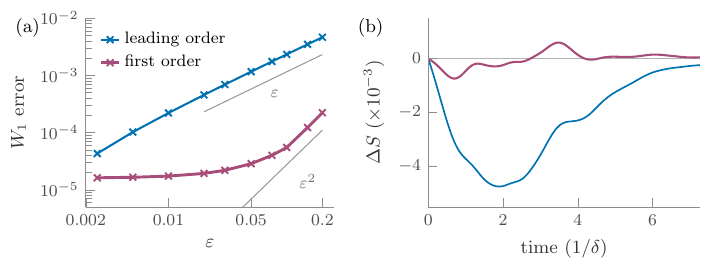}
	\caption{\textbf{For the ant foraging model the corrected PDMP is closer to the exact chain in both the food level and the stays.}
	The leading-order (blue) and corrected (purple) PDMPs are compared with the exact chain, with the parameters of \cref{fig:ants}.
	(a) Wasserstein distance $W_1$ between the exact and PDMP laws of the food level $u$ against $\varepsilon$, with guides of slope 1 and 2.
	(b) Smoothed error $\Delta S = S_{\mathrm{model}} - S_{\mathrm{exact}}$ of the survival of stays at A at $\varepsilon = 0.1$, with the PDMP paths merged as in \cref{sec:ants-residence}.}
	\label{fig:ants-errors}
\end{figure}

\Cref{fig:ants-errors} compares the leading-order and corrected PDMPs with the exact chain.
The Wasserstein error of the density of the food level falls like $\varepsilon$ at leading order and like $\varepsilon^2$ at first order, until it reaches a floor near $2\times 10^{-5}$ set by the food steps of the exact chain at $M = 8000$.
At $\varepsilon = 0.1$ the merged leading-order PDMP underestimates the survival of a stay by up to about $5\times 10^{-3}$, and the merged corrected PDMP stays within $10^{-3}$ of the exact survival.

\end{document}